\documentclass[twocolumn, eqsecnum, prd, superscriptaddress]{revtex4}

\usepackage[T1]{fontenc}
\usepackage{amsmath,amssymb}
\usepackage{amsthm}
\usepackage{appendix}
\usepackage{comment}
\usepackage[dvipdfmx]{graphicx}
\usepackage{grffile}
\usepackage{mathrsfs}
\usepackage{bm}
\usepackage{url}
\usepackage[usenames]{color}
\usepackage{longtable}
\usepackage{hyperref}

\bibstyle{apsrev4.bst}

\newcommand{\Sub}[2]{{#1}_\mathrm{#2}}
\newcommand{\Super}[2]{{#1}^\mathrm{#2}}

\newcommand{\ve}[1]{\bm{{#1}}}

\begin{document}
                                   
\title{\bf Use of 
Excess Power Method and Convolutional Neural Network in All-Sky Search for Continuous Gravitational Waves}
\date{\today}

\author{Takahiro S. Yamamoto}
\affiliation{Department of Physics, Kyoto University, Kyoto 606-8502, Japan}
\author{Takahiro Tanaka}
\affiliation{Department of Physics, Kyoto University, Kyoto 606-8502, Japan}
\affiliation{Center for Gravitational Physics, Yukawa Institute for Theoretical Physics, Kyoto University, Kyoto 606-8502, Japan}
\begin{abstract}
The signal of continuous gravitational waves has a longer duration than the observation period. Even if the waveform in the source frame is monochromatic, we will observe the waveform with modulated frequencies due to the motion of the detector. If the source location is unknown, a lot of templates having different sky positions are required to demodulate the frequency, and the required huge computational cost restricts the applicable parameter region of coherent search.
In this work, we propose and examine a new method to 
select candidates, which reduces the cost of coherent search by following-up only the selected candidates. 
As a first step, we consider an idealized situation in which 
only a single-detector having 100\% duty cycle is available 
and its detector noise is approximated by the stationary Gaussian noise.
Also, we assume the signal has no spindown and the polarization angle, the inclination angle, and the initial phase are fixed to be $\psi=0$, $\cos\iota=1$, and $\phi_0=0$, and they are treated as known parameters.
We combine several methods: 1) the short-time Fourier transform with the re-sampled data such that the Earth motion for the source is canceled 
in some reference direction, 
2) the excess power search in the Fourier transform of the time series 
obtained by picking up the amplitude in a particular frequency bin from the 
short-time Fourier transform data, and 
3) the deep learning method to further constrain the source sky position. 
The computational cost and the detection probability are estimated.
The injection test is carried out to check the validity of the detection probability.
We find that our method is worthy of further study for analyzing $O(10^7)$sec strain data.
\end{abstract}

\maketitle

\section{Introduction}

Advanced LIGO and Advanced Virgo detected the first event of gravitational waves from a binary black hole merger in 2015 \cite{gw150914}.
After the three observation runs, a lot of binary coalescence events are found \cite{gwtc1, gwtc2}.
In addition to Advanced LIGO and Advanced Virgo, KAGRA \cite{KAGRA} and LIGO India \cite{LIGOIndia} are planning to join the gravitational wave detector network \cite{FutureObs}.
The gravitational wave astronomy is expected to get fruitful results for improving our understanding of the astronomical properties of compact objects~\cite{GW170817astro, GW190521astro, GWTC-2population}, the true nature of gravity~\cite{Will2014, testofGRw/GWTC-1, testofGRw/GWTC-2}, the origin of the Universe~\cite{Maggiore2000} and so on (see~\cite{Sathyaprakash2009} for a review).

All gravitational wave signals which are detected so far have duration $O(10^{0-2})$ sec, which is much shorter than the observation period.
By contrast, we also expect gravitational waves which last longer than the observation period.
Such long-lived gravitational waves are called \textit{continuous gravitational waves} (see \cite{MaggioreBook, CreightonAndersonBook} as textbooks).
Continuous gravitational waves are defined by the following 
three properties: 1) small change rate of the amplitude, 2) almost constant fundamental frequency, and 3) duration longer than the observation period.
Rotating anisotropic neutron stars are typical candidate 
sources of continuous gravitational waves.
In addition, there are several exotic objects proposed as 
possible candidates of the sources of continuous gravitational waves (\cite{Zhu2020characterizing, Arvanitaki2015discovering, Brito2017gravitational}).

Continuous gravitational waves are modeled by simpler waveforms than those of coalescing binaries.
The parameters characterizing a typical waveform are the amplitude, the initial frequency, and the frequency derivatives with time.
Although the waveform generated by the source is analytically simple, 
the effect of the detector's motion makes the data analysis for continuous gravitational waves challenging.
The detector's motion causes the modulation in the frequency, and resulting in the dispersion of the power in the frequency domain.
If the source location is a priori known by electromagnetic observations, the modulation can be removed precisely enough.
By contrast, for the unknown target search, we need to correlate the data with a tremendous amount of templates to cover the unknown source location on the sky.
This severely restricts the applicability of 
the all-sky coherent search to strain data of long durations.
Therefore, semi-coherent methods, in which the strain data is divided into a set of segments and statistics calculated for respective segments are summed up appropriately, are often used.
Various semi-coherent methods (e.g. Time-Domain $\mathcal{F}$-statistic \cite{Jaranowski1998data}, SkyHough \cite{Krishnan2004Hough}, FrequencyHough \cite{Astone2014method}) were proposed so far, and they are actually 
used to analyze LIGO and Virgo's data.
Despite tremendous efforts, up to now no continuous gravitational wave event is detected \cite{AllSkySearchLVO1withLowerFreq, AllSkySearchLVO1withHigherFreq, AllSkySearchLVO2}.

As another trend of the research, the deep learning method is introduced to the field of  gravitational wave data analysis.
After the pioneering work done by George \& Huerta \cite{George2018deep}, 
there are many proposals to use deep learning for wide purposes, e.g., parameter estimation for binary coalescence \cite{gabbard2020bayesian, alvin2020learning, comparisonpaper, yamamoto2020use},
noise classification \cite{george2018classification}
and waveform modeling \cite{chua2020rapid}.
As for applications to the search for continuous gravitational waves, several groups already proposed deep learning methods.
Dreissigacker \textit{et al.}~\cite{Dreissigacker2019deep, Dreissigacker2020deep} applied neural networks to all-sky searches of signals with the duration $10^5$ sec and $10^6$ sec.
They used Fourier transformed strain as inputs.
Their methods can be applied to the signal located in broad frequency bands and to the case of multiple detectors and realistic noise.
Also, it is shown that the synergies between the deep learning and standard methods or other machine learning techniques are also powerful
\cite{Morawski2020convolutional, Beheshtipour2020deep}.

In this paper, we propose a new method designed for detecting monochromatic waves, 
combining several transformations and the deep learning method.
In Sec.~\ref{sec:waveformmodel}, the waveform model and some assumptions are introduced.
The coherent matched filtering and the time resampling technique are briefly reviewed in Sec.~\ref{sec:standardmethod}.
In Sec.~\ref{sec:ourmethod}, we explain our strategy that combines several traditional methods such as the resampling, the short-time Fourier transform, and the excess power search with the deep learning method.
We show the results of the assessment of the performance of our new method in Sec.~\ref{sec:results}.  Sec.~\ref{sec:conclusion} is devoted to the conclusion.


\section{Waveform model}
\label{sec:waveformmodel}

We consider a monochromatic gravitational wave. We denote by $f_\mathrm{gw}$ its frequency constant in time.
With the assumption that the source is at rest with respect to the solar system barycenter (SSB), a complex-valued gravitational waveform in the 
source frame $\Super{h}{source}(\tau)$ will be simply written as
\begin{equation}
\Super{h}{source}(\tau) = h_0 e^{2\pi i f_\mathrm{gw} \tau + i\phi_0}\,,
\label{MonochromaticSource}
\end{equation}
where $\tau$ is called \textit{SSB time}, $h_0$ and $\phi_0$ are the amplitude and the initial phase, respectively.
In this work, for simplicity, we assume that
\begin{equation}
    \phi_0 = 0\,,
    \label{assumption phi0}
\end{equation}
and regard it as a known parameter.
The phase of a gravitational wave is modulated due to the detector motion and the modulation depends on the source location.
The normal vector pointing from the Earth's center to the sky position specified by a right ascension $\alpha$ and a declination angle $\delta$ is defined by
\begin{equation}
    \ve{n}(\alpha, \delta) =
    \left(\begin{matrix}
    1 & 0 & 0 \\
    0 & \cos\epsilon & \sin\epsilon \\
    0 & -\sin\epsilon & \cos\epsilon 
    \end{matrix}\right)
    \left(\begin{matrix}
    \cos\alpha\cos\delta \\ \sin\alpha\cos\delta \\ \sin\delta
    \end{matrix}\right)\,,
\label{sourcedirection}
\end{equation}
with the tilt angle between the Earth's rotation axis and the orbital angular momentum $\epsilon$.
Here, we work in the SSB frame, in which the $z$-axis is along the Earth's orbital angular momentum and the $x$-axis points towards the vernal equinox.
Defining the detector time $t$ so as to satisfy 
\begin{equation}
    \tau = t + \frac{\ve{r}(t)\cdot \ve{n}(\Sub{\alpha}{s}, \Sub{\delta}{s})}{c}\,,
\label{eq:SSBtime_and_detectortime}
\end{equation}
we obtain the waveform in the detector frame
\begin{equation}
    h(t) := h_0 e^{i\Phi(t)}\,,
\end{equation}
with
\begin{equation}
\Phi(t) = 2\pi f_\mathrm{gw} t + 2\pi f_\mathrm{gw} \frac{\ve{r}(t)\cdot \ve{n}(\alpha_\mathrm{s}, \delta_\mathrm{s})}{c}\,.
\label{eq:modulatedphase}
\end{equation}
A subscript ``s'' indicates the quantity related to the gravitational wave source.
Namely, $(\alpha_\mathrm{s}, \delta_\mathrm{s})$ means the sky position of the source.
In the following, we use the notation $\ve{n}_\mathrm{s} := \ve{n}(\alpha_\mathrm{s}, \delta_\mathrm{s})$.

For the modeling of the detector motion, we adopt a little simplification, which we believe will not affect our main result. 
We assume that the position vector of the detector can be written by a sum of the Earth's rotation part $\ve{r}_\oplus (t)$, and the Earth's orbital motion part $\ve{r}_\odot (t)$.
The Earth is assumed to take a circular orbit on $xy$-plane.
Then, we can write $\ve{r}_\odot(t)$ as
\begin{equation}
\ve{r}_\odot(t) = R_\mathrm{ES}
\left(\begin{matrix}
\cos (\varphi_\odot + \Omega_\odot t) \\ \sin (\varphi_\odot + \Omega_\odot t) \\
0
\end{matrix}\right)\,,
\label{solarmotion}
\end{equation}
where $R_\mathrm{ES}$, $\Omega_\odot$ and  $\varphi_\odot$ are 
the distance between the Earth and the Sun, the angular velocity of the orbital motion and the initial phase, respectively.
The detector motion due to the Earth's rotation can be described as
\begin{equation}
\ve{r}_\oplus(t)
=R_\mathrm{E}
\left(\begin{matrix}
1 & 0 & 0 \\
0 & \cos\epsilon & \sin\epsilon \\
0 & -\sin\epsilon & \cos\epsilon 
\end{matrix}\right)
\left(\begin{matrix}
\cos\lambda\cos (\varphi_\oplus + \Omega_\oplus t) \\ \cos\lambda\sin(\varphi_\oplus + \Omega_\oplus t) \\ \sin\lambda
\end{matrix}\right)\,,
\label{eq:detloc}
\end{equation}
where $R_\mathrm{E}$, $\lambda$, $\Omega_\oplus$ and $\varphi_\oplus$ are 
the radius of the Earth, the latitude of the detector, 
the angular velocity of the Earth's rotation
and the initial phase, respectively.
The modulated phase $\Phi(t)$ can be decomposed into
\begin{equation}
    \Phi(t) = 2\pi \Sub{f}{gw} t + \Phi_\oplus(t) + \Phi_\odot(t),
\end{equation}
where
\begin{align}
    \Phi_\oplus(t) &= 2\pi \Sub{f}{gw} \frac{\ve{r}_\oplus(t) \cdot \Sub{\ve{n}}{s}}{c}, \\
    \Phi_\odot(t) &= 2\pi \Sub{f}{gw} \frac{\ve{r}_\odot(t) \cdot \Sub{\ve{n}}{s}}{c}.
\end{align}

Finally, we take into account 
the amplitude modulation due to the detector's motion, 
which can be described by the antenna pattern function.
In this work, the polarization angle and the inclination angle are, respectively, assumed to be
\begin{equation}
    \psi=0\,, \qquad \cos\iota=1\,,
    \label{assumption psi cosi}
\end{equation}
and, similarly to $\phi_0$, they are treated as known parameters.
Then, the gravitational wave to be observed by a detector can be written as 
\begin{equation}
\Sub{h}{obs}(t) = G(t) h(t) + G^\ast(t) h^\ast \,,
\label{eq:hdet}
\end{equation}
with
\begin{equation}
    G(t) := \frac{F_+(t) + iF_\times(t)}{2}\,.
\end{equation}
The definitions of $F_+(t)$ and $F_\times(t)$ are the same as those used in Jaranowski \textit{et al.,} \cite{Jaranowski1998data}.
In this work, the antenna pattern function of LIGO Hanford is employed.
The strain data is written as
\begin{equation}
    s(t) = \Sub{h}{obs}(t) + n(t)\,,
\label{eq:s(t)}
\end{equation}
where $n(t)$ is the detector noise.
We assume that 
the strain data from the detector has no gaps in time 
and the detector noise is stationary and Gaussian.

\section{Coherent search method}
\label{sec:standardmethod}

Before explaining our method, we briefly review the coherent search method and the time resampling technique \cite{Jaranowski1998data}.

If the expected waveforms can be modeled precisely and the noise is Gaussian, 
the matched filtering is the optimal method for the detection and the 
parameter estimation, besides the computational cost.
The noise weighted inner product is defined by
\begin{equation}
(a|b) := 4 \mathrm{Re} \left[ \int^\infty_0 df\ \frac{\tilde{a}(f) \tilde{b}^\ast(f)}{\Sub{S}{n}(f)} \right]\,,
\label{eq:InnerProduct}
\end{equation}
where $\Sub{S}{n}(f)$ is the power spectral density of the detector noise.
A signal-to-noise ratio (SNR) can be calculated with the inner product between a strain $s(t)$ and a template $\Sub{h}{temp}(t)$ as
\begin{equation}
\Sub{\rho}{MF} := \frac{(s|\Sub{h}{temp})}{\sqrt{(\Sub{h}{temp}|\Sub{h}{temp})}}\,.
\label{eq:rho_MF}
\end{equation}
Theoretically predicted waveforms $\Sub{h}{temp}(t)$ have various parameters characterizing the source properties and the geometrical information.
A set of waveforms having different parameters is called a template bank.
For each template in a template bank, we can assign the value of SNR calculated by Eq.~\eqref{eq:rho_MF}.
If the maximum value of SNR in the template set exceeds a threshold value, 
it is a sign that an actual signal may exist and the parameter inference is also obtained from the distribution of SNR.

Due to a long duration and a narrow frequency band of continuous gravitational waves, 
the inner product~\eqref{eq:InnerProduct} can be recast into the time-domain 
expression as
\begin{equation}
(a|b) \simeq \frac{2}{\Sub{S}{n}(\Sub{f}{gw})} \mathrm{Re} \left[ \int^{\Sub{T}{obs}}_0 dt\ a(t) b^\ast(t) \right]\,,
\label{eq:timedomain_innerproduct}
\end{equation}
where $\Sub{T}{obs}$ is the observation time.
For a monochromatic source, the waveform can be modeled by Eq.~\eqref{MonochromaticSource}.
The modulation due to the detector motion is only in the phase of the waveform.
The time resampling technique nullifies the phase modulation by redefining the time coordinate.
If the position of the source is a priori known, the exact relation \eqref{eq:SSBtime_and_detectortime} can be obtained.
Therefore, the phase modulation can be completely removed and the monochromatic waveform is applicable to the time-domain matched filtering \eqref{eq:timedomain_innerproduct}.
Calculation of the inner product \eqref{eq:timedomain_innerproduct} between the resampled signal and a monochromatic waveform is equivalent to the Fourier transform.
Thus, the fast algorithm (i.e., Fast Fourier transform) can be employed to rapidly search the gravitational wave frequency, $\Sub{f}{gw}$.

When the source location is unknown, 
we need to search all-sky by placing a set of grid points $\{\Sub{\ve{n}}{g}^{(i)}\}_{i=1}^{\Sub{N}{grid}}$ to keep the maximum loss of SNR 
within the acceptable range.
For each grid point, we carry out the Fourier transform 
after the transformation
\begin{equation}
    \zeta := t + \frac{\ve{r}(t)\cdot \Sub{\ve{n}}{g}}{c}\,. 
\end{equation}
Here, we omit the superscript $(i)$, for brevity.
The necessary number of grid points $\Sub{N}{grid}$ can be estimated by the angular resolution of gravitational wave sources.
The angular resolution of gravitational wave sources can be roughly 
given by the ratio between the wavelength of the gravitational wave 
and the diameter of the Earth's orbit, i.e.,
\begin{equation}
\Sub{(\delta\theta)}{coh} \sim \frac{\lambda_\mathrm{gw}}{2R_\mathrm{ES}} \sim 10^{-5} \text{ [rad]}\,.
\label{eq:resolution_coh}
\end{equation}
Here, we adopt 100Hz as the fiducial value for $c\Sub{\lambda}{gw}^{-1}$. 
Thus, the required number of grid points is, at least, 
\begin{equation}
N_\mathrm{grids} \sim \frac{4\pi}{\Sub{(\delta\theta)}{coh}^2} \sim 1.3 \times 10^{11}\,.
\end{equation}
The time resampling and the Fourier transform are applied to each grid point.
The number of floating point operations required for carrying out FFT is $\sim 1.7 \times 10^{12}$ per grid point,
with the signal of a duration $10^7$ sec and a sampling frequency 1024 Hz.
Even if we have a 1PFlops machine, the computational time becomes $2.2 \times 10^8$ sec, which is longer than the signal duration.
For this reason, the coherent all-sky search is not realistic even for monochromatic sources yet.

\section{Our Method}
\label{sec:ourmethod}

\subsection{Subtracting the effect due to the Earth's rotation}

As stated in Sec.~\ref{sec:standardmethod}, the time resampling technique can demodulate the phase, but complete demodulation is not available because of the limitation of computational resources.
In our work, the time resampling technique is employed to eliminate only the effect caused by the Earth's diurnal rotation, $\Phi_\oplus(t)$.
Assuming a representative grid point $\ve{n}_\mathrm{g}$,
we can rewrite the phase \eqref{eq:modulatedphase} as
\begin{align}
    \Phi(t) &= 2\pi \Sub{f}{gw} t + \Phi_\oplus(t) + \Phi_\odot(t) \notag \\
    &= 2\pi \Sub{f}{gw} \zeta + \delta\Phi_\oplus(t) + \delta\Phi_\odot(t)\,,
    \label{eq:resampledphase}
\end{align}
where
\begin{align}
\delta\Phi_\oplus(t) &:= 2\pi f_\mathrm{gw} \frac{\ve{r}_\oplus(t) \cdot \Delta\ve{n}}{c}\,,
\label{eq:dPhiEarth} \\
\delta \Phi_\odot(t) &:= 2\pi f_\mathrm{gw} \frac{\ve{r}_\odot(t) \cdot \Delta\ve{n}}{c}\,,
\label{eq:dPhiSun}
\end{align}
and $\Delta \ve{n}:=\ve{n}_\mathrm{s} - \ve{n}_\mathrm{g}$.
Since the residual phase varies with time, we will place grid points so that the amplitude of the residual phase in the worst case, i.e.,
$$
\min_{\ve{n}_\mathrm{g}}\max_t |\delta\Phi_\oplus(t)|
$$ 
becomes smaller than a threshold $\delta\Phi_\epsilon$ for any source direction $\ve{n}_\mathrm{s}$ within the area covered by the grid point $\Sub{\ve{n}}{g}$. 
To optimize the grid placement, we employ the method proposed in Ref.~\cite{Nakano2003effective}.
The residual phase $\delta\Phi_\oplus$ is expanded up to the first order of $\Delta\alpha:=\alpha_\mathrm{s}  - \alpha_\mathrm{g}$ and $\Delta\delta := \delta_\mathrm{s} - \delta_\mathrm{g}$.
Then, we get
\begin{align}
\delta \Phi_\oplus \simeq &\frac{2\pi \Sub{f}{gw}}{c} R_\text{E} \cos\lambda \left\{ - \Delta\delta \sin\delta_\mathrm{g} \cos(\alpha_\mathrm{g} - \varphi_\oplus - \Omega_\oplus t) \right. \notag \\
&\left. - \Delta\alpha \cos\delta_\mathrm{g} \sin(\alpha_\mathrm{g} - \varphi_\oplus - \Omega_\oplus t) \right\}\,.
\end{align}
Here, the constant term is neglected because it degenerates with the initial phase $\phi_0$.
The maximum value of the residual phase is
\begin{align}
\max_t |\delta \Phi_\oplus| = &\frac{2\pi \Sub{f}{gw}}{c} R_\text{E} |\cos\lambda| \notag \\
&\times \sqrt{ (\Delta\delta)^2 \sin^2\Sub{\delta}{g}+ (\Delta\alpha)^2 \cos^2\Sub{\delta}{g} }\,.
\label{eq:max_dPhiEarth}
\end{align}
The grid points are to be determined to satisfy 
$\max_t |\delta\Phi_\oplus| < \delta\Phi_\epsilon$ 
for any source direction.

Because the residual phase \eqref{eq:max_dPhiEarth} is symmetric under the transformation $\Sub{\delta}{g} \to -\Sub{\delta}{g}$,
the placement of grids on the negative $\delta$ side can be generated by inverting the sign of the grids on the positive $\delta$ side.
Therefore, we focus on the case with $0\leq \delta \leq \pi/2$.

Since the residual phase depends only on $\delta$ at $\delta_\mathrm{g} = \pi/2$,
a single template can cover the neighbor of $\delta = \pi/2$.
In fact, at $\delta = \pi/2$, Eq.~\eqref{eq:max_dPhiEarth} becomes
\begin{equation}
\max_t |\delta \Phi_\oplus| = \frac{2\pi \Sub{f}{gw}}{c} R_\text{E} |\Delta\delta| \cos\lambda \,.
\end{equation}
Therefore, the condition $\max_t |\delta\Phi_\oplus| \leq \delta\Phi_\epsilon$ gives the lower bound of $\delta_1$ such that the region $\delta_1 \leq \delta \leq \pi/2$ can be covered by a signle patch represented by $\{(\Sub{\alpha}{g}, \Sub{\delta}{g}) = (0, \pi/2) \}$, 
to find
\begin{equation}
\delta_1 := \frac{\pi}{2} - \delta\Phi_\epsilon \times \frac{c}{2\pi \Sub{f}{gw}} \frac{1}{R_\text{E} \cos\lambda}\,.
\end{equation}

\begin{figure}[ht]
\includegraphics[width=8.0cm]{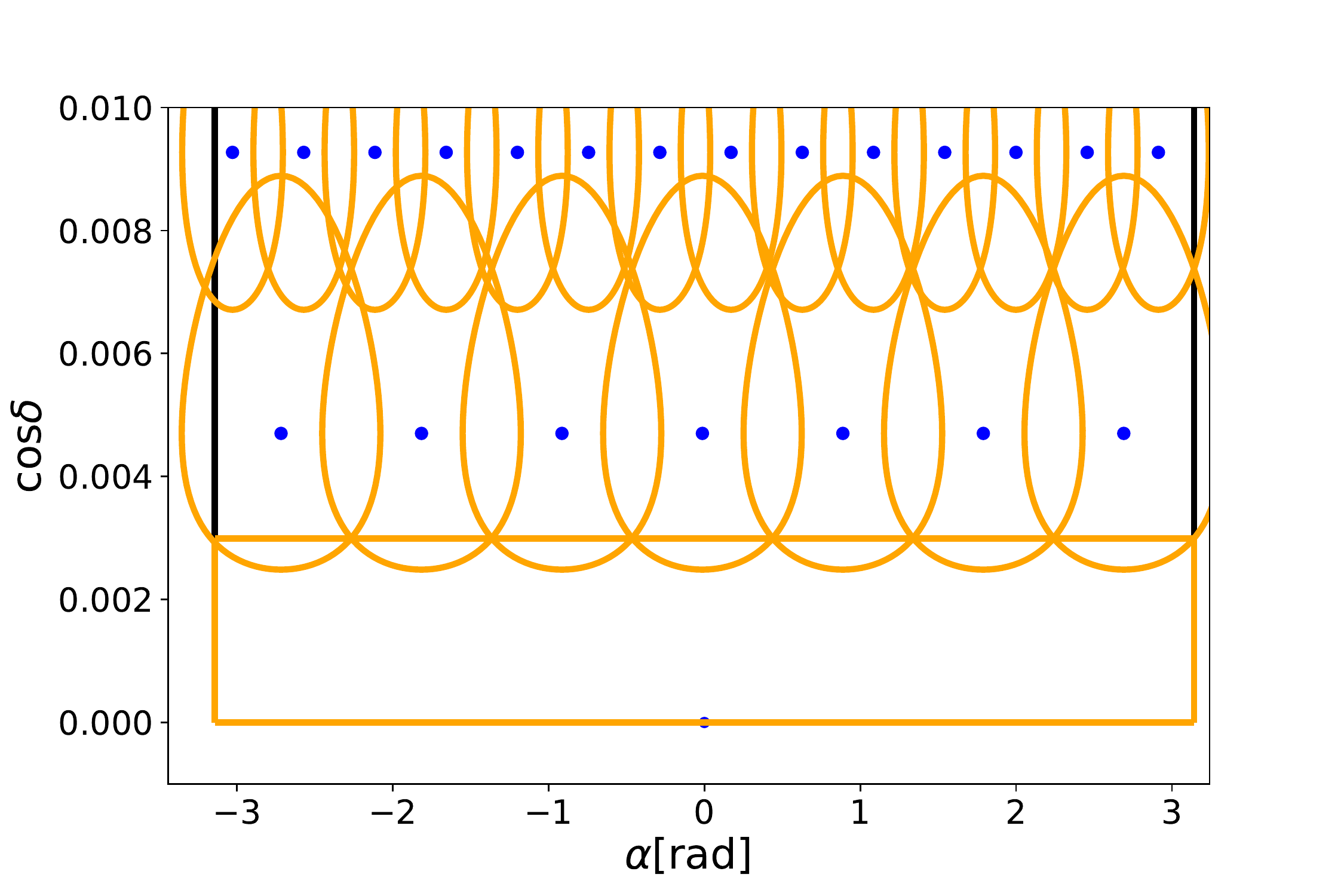}
\caption{\label{fig:patches}
Grid point placement on a fraction of $(\alpha, \cos\delta)$-plane.
Blue dots are grid points and orange contours show the $\max_t |\delta \Phi_\oplus(t)|=\delta\Phi_\epsilon$ contours for each grid point.
The region $\{(\alpha, \delta) | \delta<\delta_1\}$ is covered by a single template $(\alpha_\mathrm{g}, \delta_\mathrm{g}) = (0, \pi/2)$ and the shape of the patch is square on this plane.}
\end{figure}

Plural patches are necessary to cover the strip of a constant $\delta$ in the other range. 
We introduce a 2-dimensional metric corresponding to the residual phase \eqref{eq:max_dPhiEarth}, 
\begin{equation}
d\sigma^2 = \cos^2\delta d\alpha^2 + \sin^2\delta d\delta^2\,.
\label{metric}
\end{equation}
In general, a metric in a 2-dimensional manifold can be transformed into a conformally flat metric by an appropriate coordinate transformation.
When the space is conformally flat, the curve of a small constant distance measured from an arbitrary chosen point can be approximated by a circle.
Therefore, a template spacing in the 2-dimensional parameter space becomes relatively easy.
By defining new variables $X := \alpha$ and $Y := -\log |\cos\delta|$, the metric can be transformed into
\begin{equation}
	d\sigma^2 = e^{-2Y} (dX^2 + dY^2)\,.
\end{equation}
Along with \cite{Nakano2003effective}, we can construct the sky patches covering the half-sky region with $0\leq \delta \leq \delta_1$.
Figure~\ref{fig:patches} shows a part of grid points constructed under the condition
\begin{equation}
    \delta\Phi_\epsilon = 0.058\,,
\end{equation}
which we adopt throughout this paper.
The total number of grid points to cover the whole sky 
is 
\begin{equation}
  N_{\rm grid}=352,436\,, 
\end{equation}
for $f_\mathrm{gw}$ = 100Hz.

\subsection{Modeling the effect due to the Earth's orbital motion}
\label{sec:modeling}

As we choose $\delta\Phi_\epsilon$ to be sufficiently small, we
neglect $\delta\Phi_\oplus$ in the following discussion. 
Then, after subtracting the phase modulation due to the Earth's rotation,
the phase of the gravitational wave \eqref{eq:resampledphase} becomes 
\begin{equation}
    \Phi(t) = 2\pi \Sub{f}{gw} \zeta + \delta\Phi_\odot(t).
    \label{eq:phase_with_dPhiSun}
\end{equation}
We apply the short-time Fourier transform (STFT) to the time-resampled strain,
\begin{equation}
    s(\zeta) = \Sub{h}{obs}(\zeta) + n(\zeta).
    \label{eq:timeresampledstrain}
\end{equation}
In the rest of the paper, we treat only the time-resampled data.
Therefore, without confusion,
the time-resampled data in Eq.~\eqref{eq:timeresampledstrain} can be denoted by the same character as the original one.
The strain is divided into $\Sub{N}{seg}$ segments having the duration $\Sub{T}{seg}$ and their start times are denoted by $\zeta_j := j T_\mathrm{slide},\ (j=0,1,\cdots,\Sub{N}{seg}-1)$.
$T_\mathrm{slide}$ is not necessary to be equal to $T_\mathrm{seg}$. 
The output of STFT with the window function $w(\zeta)$ is defined by
\begin{equation}
    s^\mathrm{STFT}_{j,k} = h^\mathrm{STFT}_{j,k} + n^\mathrm{STFT}_{j,k}\,,
    \label{eq:s_STFT}
\end{equation}
where
\begin{equation}
h^\mathrm{STFT}_{j,k} = \frac{1}{T_\mathrm{seg}} \int^{\zeta_j + T_\mathrm{seg}}_{\zeta_j} d\zeta'\ w(\zeta' - \zeta_j) \Sub{h}{obs}(\zeta') e^{-2\pi i f_k \zeta'}\,,
\label{eq:h_STFT}
\end{equation}
\begin{equation}
    n^\mathrm{STFT}_{j,k} = \frac{1}{T_\mathrm{seg}} \int^{\zeta_j + T_\mathrm{seg}}_{\zeta_j} d\zeta'\ w(\zeta' - \zeta_j) n(\zeta') e^{-2\pi i f_k \zeta'}
    \label{eq:n_STFT}\,,
\end{equation}
and $f_k := k\Delta f = k/T_\mathrm{seg}$ is the frequency of the $k$-th element of STFT.
Let us focus on the positive frequency modes, i.e., $f_k>0$.
Then, the second term of Eq.~\eqref{eq:hdet} can be neglected and Eq.~\eqref{eq:h_STFT} can be approximated by
\begin{align}
    \Super{h}{STFT}_{j,k} \simeq& \frac{1}{\Sub{T}{seg}} \int^{\zeta_j + \Sub{T}{seg}}_{\zeta_j} d\zeta'\ \bigg\{w(\zeta' - \zeta_j) \notag\\
    &\times G(t(\zeta')) e^{2\pi i \delta f_k \zeta'} e^{i\delta\Phi_\odot(\zeta')} \bigg\}.
    \label{eq:hSTFT_interm}
\end{align}
with $\delta f_k := \Sub{f}{gw} - f_k$.
In the expression of $G(t(\zeta))$, the SSB time $\zeta$ appears only through the combination $\Omega_\oplus t(\zeta)$.
The difference between $\Omega_\oplus t(\zeta)$ and $\Omega_\oplus \zeta$ is 
negligiblly small. 
Therefore, in Eq.~\eqref{eq:hSTFT_interm}, $G(t(\zeta))$ can be replaced by $G(\zeta)$.
The duration $\Sub{T}{seg}$ is chosen so that $G(t(\zeta))$ can be approximated by a constant in each segment.
With this choice of $\Sub{T}{seg}$, the factor $e^{i\delta\Phi_\odot(t)}$ also can be seen as a constant in each segment because it varies slower than the antenna pattern function.
Therefore, Eq.~\eqref{eq:hSTFT_interm} can be approximated by
\begin{equation}
h^\mathrm{STFT}_{j,k} \simeq h_0 e^{i\delta\Phi_\odot(\zeta_j)} G(\zeta_j) W_k(\zeta_j)\,,
\label{eq:hSTFT}
\end{equation}
where
\begin{equation}
W_k(\zeta_j) := \frac{1}{T_\mathrm{seg}} \int^{\zeta_j + T_\mathrm{seg}}_{\zeta} d\zeta'\ w(\zeta' - \zeta_j) e^{2\pi i \delta f_k \zeta' }\,.
\label{eq:W}
\end{equation}
In this work, we use the tukey window,  \begin{align}
&w(\zeta) = \notag\\
&\begin{cases}
\frac{1}{2} - \frac{1}{2}\cos \left(\frac{2\pi \zeta}{\alpha T_\mathrm{seg}} \right)\,, & 
\left(0 \leq \frac{\zeta}{T_\mathrm{seg}} < \frac{\alpha}{2}\right)\,, \\
1\,, & \left(\frac{\alpha}{2} \leq \frac{\zeta}{T_\mathrm{seg}} \leq 1-\frac{\alpha}{2}\right)\,, \\
\frac{1}{2} - \frac{1}{2} \cos \left( \frac{2\pi (T_\mathrm{seg} - \zeta)}{\alpha T_\mathrm{seg}} \right)\,, & 
\left(1-\frac{\alpha}{2} < \frac{\zeta}{T_\mathrm{seg}} \leq 1\right)\, .
\end{cases}
\end{align}
We set the parameter $\alpha$ to 0.125.
With $\beta_k := T_\mathrm{seg} \delta f_k$, Eq.~\eqref{eq:W} can be calculated as
\begin{align}
W_k(\zeta_j)
=&\ e^{2\pi i \delta f_k \zeta_j} \notag\\
&\times \frac{ (1+e^{i\pi \alpha \beta_k}) (1- e^{2\pi i \beta_k (1-\alpha/2)} )}{4\pi i \beta_k (\alpha^2\beta_k^2 - 1)}\,.
\end{align}

\begin{figure}[t]
    \centering
    \includegraphics[width=8cm]{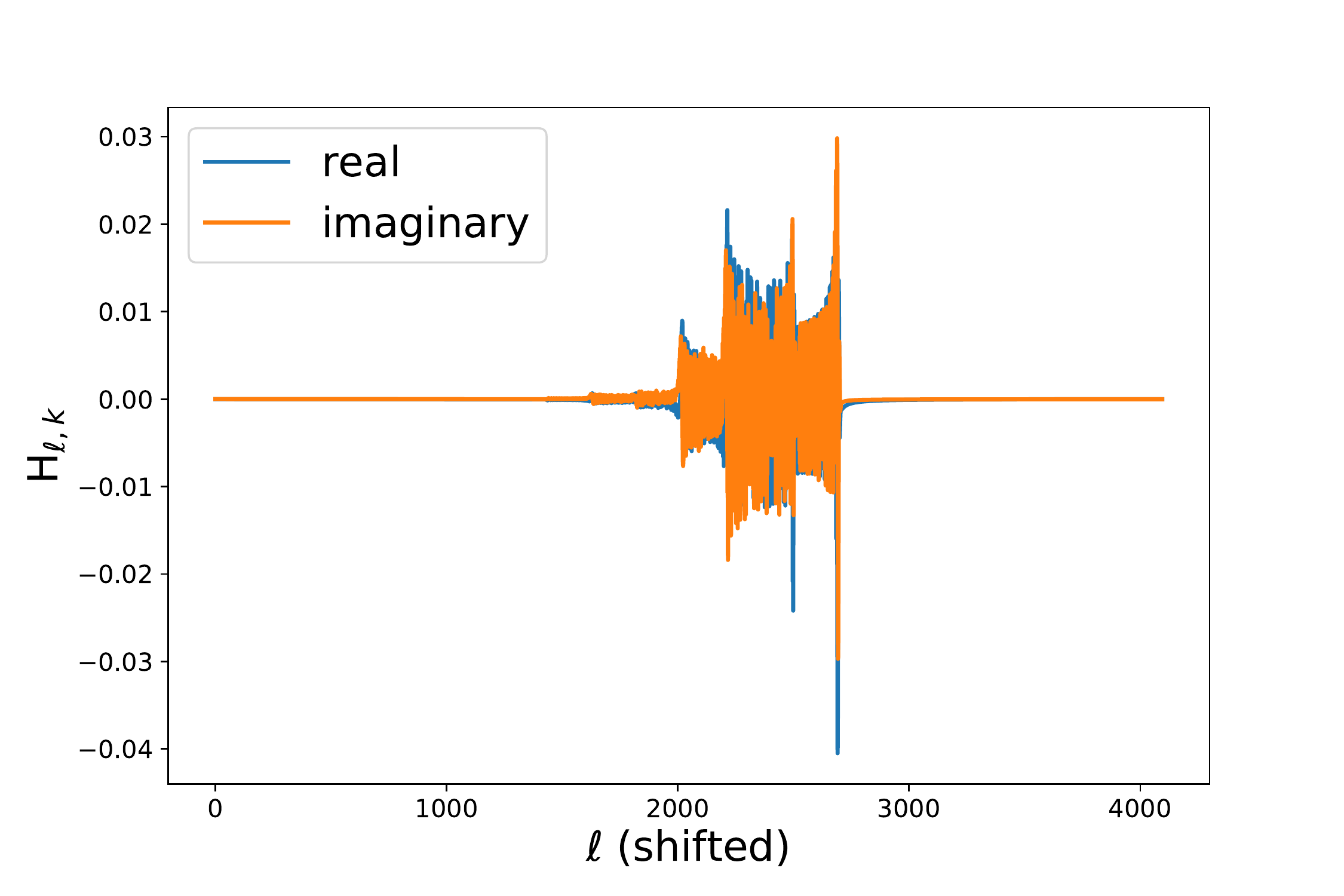}
    \caption{\label{fig:waveform}
    An example of the $\ell$-domain waveform.
    The length of the $\ell$-domain waveform is $2^{19}$. This figure is a zoom-in around the region at where the signal is localized.
    The amplitude is $h_0=1.0$.}
\end{figure}

Using the Jacobi-Anger expansion, we can expand the factor $e^{i\delta\Phi_\odot(\zeta_j)}$ that appears in Eq.~\eqref{eq:hSTFT} as
\begin{equation}
    e^{i\delta\Phi_\odot(\zeta_j)} = \sum_{\ell=-\infty}^\infty i^\ell J_\ell(X) e^{i\ell \Omega_\odot \zeta_j} e^{i\ell(\varphi_\odot - \phi_X)}
    \label{eq:expansioninBessel}
\end{equation}
where $J_\ell(z)$ is the Bessel function of the first kind and 
\begin{align}
    & X := \frac{2\pi \Sub{f}{gw} \Sub{R}{ES}}{c} \sqrt{(\Delta n_x)^2 + (\Delta n_y)^2}, \\
    & e^{i\phi_X} := \Delta n_x + i \Delta n_y.
\end{align}
Therefore, Eq.~\eqref{eq:hSTFT} can be expressed as
\begin{align}
    \Super{h}{STFT}_{j,k} &\simeq h_0 G(\zeta_j) W_k(0) e^{2\pi i\delta f_k \zeta_j} \notag\\
    &\times \sum_{\ell=-\infty}^\infty i^\ell  J_{\ell}(X) e^{i\ell(\varphi_\odot - \phi_X)} e^{i\ell\Omega_\odot \zeta_j}\,.
\end{align}
The Fourier transform of $\Super{h}{STFT}_{j,k}$ with a fixed integer $k$ is defined by
\begin{equation}
\mathsf{H}_{\ell,k} := \frac{1}{\Sub{N}{seg}} \sum_{j=0}^{\Sub{N}{seg}-1} \Super{h}{STFT}_{j,k} e^{-2\pi i j\ell/ \Sub{N}{seg}}.
\label{eq:def_of_H}
\end{equation}
We refer to $\mathsf{H}_{\ell,k}$ as the \textit{$\ell$-domain signal}.
To understand the pattern hidden in $\mathsf{H}_{j,k}$,
we set aside the factor $G(\zeta_j)$ for a while.
Then, Eq.~\eqref{eq:def_of_H} can be estimated as
\begin{equation}
\mathsf{H}_{\ell,k} \sim h_0 W_k(0) i^{\ell'} J_{\ell'}(X) e^{i\ell' (\varphi_\odot - \phi_X)},
\label{eq:ell_domain_signal}
\end{equation}
with $\ell' \sim \ell + 2\pi \Omega_{\odot}^{-1} \delta f_k$.
Because of the fact that $J_\ell(z) \simeq 0$ for $|\ell| \gtrsim |z|$ and $X\lesssim O(10^3)$ for $\Sub{f}{gw}=100$Hz, the signal is localized within the region where only few thousand bins in the $\ell$-domain.
Putting back the antenna pattern $G(\zeta_j)$, we expect that $\ell$-domain signals lose their amplitude and their localizations become worse than those for the idealized cases.
Figure~\ref{fig:waveform} shows an example of the $\ell$-domain signal.
%

\subsection{Excess power method for finding candidates}
\label{sec:EPmethod}

By the method shown in the previous subsection, for every grid point $\ve{n}_\mathrm{g}$ and every frequency bin $f_k$, we obtain an $\ell$-domain strain defined by
\begin{equation}
    \mathsf{S}_{\ell,k} := \mathsf{H}_{\ell,k} + \mathsf{N}_{\ell,k},
\end{equation}
with
\begin{equation}
    \mathsf{N}_{\ell,k} := \frac{1}{\Sub{N}{seg}} \sum_{j=0}^{\Sub{N}{seg}-1} \Super{n}{STFT}_{j,k} e^{-2\pi i j\ell / \Sub{N}{seg}}.
\end{equation}
There are $\Sub{T}{obs}/\Sub{T}{seg} \sim O(10^6)$ data points in a single $\ell$-domain strain and we know that the signal in $\ell$-domain will be localized within a small region $\sim O(10^3)$.
Thus, the excess power method \cite{anderson2001excess} is useful for selecting the candidates with a minimal computational cost.
We here divide an $\ell$-domain signal into short chunks so that each chunk has the length $\delta\ell$ and neighbored segments have an overlap by $\delta\ell/2$, which is 
one of the simplest choices but not the optimal one. 
Then, we obtain $\Sub{N}{chunk/signal}=2(\Sub{N}{seg}-\delta\ell)/\delta\ell$ chunks from one $\ell$-domain signal.
The excess power statistic for the grid point $\Sub{\ve{n}}{g}$, the frequency bin $f_k$, 
and the $c$-th chunk ($c = 0,1,\dots,\Sub{N}{chunk/signal}-1$) is defined by
\begin{equation}
\mathcal{E} (\Sub{\ve{n}}{g}, f_k, c) := 4 \sum_{\ell=c\delta\ell/2}^{(c+2)\delta\ell/2-1}\frac{|\mathsf{S}_{\ell,k}|^2}{\tilde{\sigma}_k^2}\,,
\end{equation}
where
\begin{equation}
    \langle \mathsf{N}_{\ell,k} \mathsf{N}^\ast_{\ell',k} \rangle =: \frac{1}{2}\tilde{\sigma}_k^2 \delta_{\ell \ell'}\,.
    \label{<NN>}
\end{equation}
The variance of noise in $\ell$-domain, $\tilde{\sigma}_k$, is estimated as
\begin{equation}
    \tilde{\sigma}_k^2 = \frac{\Sub{S}{n}(f_k)}{\Sub{N}{seg} \Sub{T}{seg}} \times \mathcal{W}\,,
    \label{eq:sigmatilde}
\end{equation}
where $\mathcal{W}$ is the factor coming from the window function and defined by
\begin{equation}
    \mathcal{W} := \int^{0.5}_{-0.5} dx [w(x)]^2\,.
\end{equation}
The derivation of Eq.~\eqref{eq:sigmatilde} is summarized in Appendix~\ref{sec:appendix}.

We define the SNR of the excess power by
\begin{equation}
\Sub{\rho}{EP}(\Sub{\ve{n}}{g},f_k,c) := \frac{\mathcal{E}(\Sub{\ve{n}}{g},f_k,c) - \langle \mathcal{E} \rangle_\mathrm{n}}{\sigma_\mathrm{n}(\mathcal{E})}\,,
\end{equation}
where 
\begin{equation}
    \langle \mathcal{E} \rangle_\mathrm{n}=2\delta \ell\,,
\end{equation}
and 
\begin{equation}
    \sigma_\mathrm{n}(\mathcal{E}):=
\sqrt{\langle (\mathcal{E} -\langle \mathcal{E} \rangle_\mathrm{n})^2\rangle_\mathrm{n}
}=2\sqrt{\delta \ell}\,,
\end{equation}
are, respectively, the expectation value and the standard deviation of $\mathcal{E}$ 
when only noise exists.
We select the candidate set of parameter values $\{\Sub{\ve{n}}{g}, f_k, c\}$, when
\begin{equation*}
    \Sub{\rho}{EP}(\Sub{\ve{n}}{g},f_k,c) > \Sub{\hat{\rho}}{EP}
\end{equation*}
is satisfied with a threshold value $\Sub{\hat{\rho}}{EP}$.
Strictly speaking, 
since the excess power statistic $\mathcal{E}$ is the sum of $2\delta\ell$ squared 
Gaussian random variables with the variance $1/2\sqrt{\delta\ell}$, 
$\mathcal{E}$ follows a chi square distribution with the degree of freedom $2\delta\ell$.
However, since here we choose $\delta\ell$ to be large, 
the distribution of $\mathcal{E}$ can be approximated by a Gaussian distribution with the average $2\delta\ell$ and the standard deviation $2\sqrt{\delta \ell}$. 
Therefore, in the absence of gravitational wave signal, the probability distribution of $\Sub{\rho}{EP}$ is a Gaussian distribution with zero mean and unit variance.

Also in the presence of some signal, 
the excess power statistics $\Sub{\rho}{EP}$ is given by 
a sum of many statistical variables. 
Thus, the statistical distribution of $\Sub{\rho}{EP}$ can be approximated 
by the Gaussian distribution whose 
mean and standard deviation are calculated as
\begin{equation}
\Sub{\mu}{EP}(\ve{\xi})= \frac{2P_k(\ve{\xi})}{\tilde{\sigma}^2_{k} \sqrt{\delta\ell}}\,,
\label{eq:rhomean_EP}
\end{equation}
and 
\begin{equation}
\Sub{\sigma}{EP}(\ve{\xi}) = \sqrt{ 1 + \frac{4P_k(\ve{\xi})}{\tilde{\sigma}_k^2 \delta\ell}}\,,
\label{eq:rhostd_EP}
\end{equation}
Here, we define
\begin{equation}
P_k(\ve{\xi}) := \sum_{\ell} |\mathsf{H}_{\ell,k}(\ve{\xi})|^2\,,
\label{eq:power_of_H}
\end{equation}
and we define $\ve{\xi}$ as a set of parameters $\ve{\xi} := (h_0, \vec{\xi}) = (h_0, \Sub{f}{gw}, \Sub{\alpha}{s}, \Sub{\delta}{s})$.
The false alarm rate and the detection efficiency will be assessed with this Gaussian approximation.

\subsection{Neural network for localizing}
\label{sec:NN}

\subsubsection{fundamentals}

Deep learning is one of the approaches for finding features being hidden in the data (see \cite{Goodfellow2016} as a textbook).
Artificial neural networks (ANNs) are the architectures playing the central roll in deep learning.
An ANN consists of consecutive layers and each layer is formed by a lot of units (neurons).
Each layer takes inputs from the previous layer and processed data is passed to the next layer.
As a simple example, the process occurring in each layer can be written as the combination of affine transformation and a non-linear transformation, i.e.,
\begin{equation}
x^{(\ell+1)}_i = g\left( \sum_{j=1}^{N^{(\ell)}} w^{(\ell)}_{ij} x^{(\ell)}_j + b^{(\ell)} \right)\ 
(i = 1,2,\cdots, N^{(\ell+1)})\,, 
\label{MPL}
\end{equation}
where $x^{(\ell)}$ is a set of input data on the $\ell$-th layer and $g$ is a nonlinear function, which is called an activation function.
We use a ReLU function \cite{ReLU}, defined by
\begin{equation}
g(z) = \max [z, 0]\,.
\end{equation}
The parameters $w$ and $b$ are, respectively, called weights and biases.
They are tunable parameters and optimized to capture the features of data.
The process to optimize weights and biases is called training.
Frequently, the affine transformation and the non-linear transformation 
are divided into two layers, called a linear layer and 
a non-linear transformation layer, respectively.

In addition to the layers as given by Eq.~\eqref{MPL},
many variants are proposed so far.
In this work, we use also one-dimensional convolutional layers \cite{LeCun1998} and max-pooling layers \cite{MaxPooling}.
The input of a convolutional layer, denoted by $x^{c}_i$, is a set of vectors.
For example, in the case of color images, each pixel has three channels corresponding to three primary colors of light.
Therefore, the input data is a set of three two-dimensional arrays.
The discrete convolution, which is represented as
\begin{equation}
o^{c'}_{i} = \sum_{c=0}^{C-1} \sum_{k=0}^{K-1} f^{c,c'}_k x^c_{i+k} + b^{c'}\,,
\end{equation}
is calculated in a convolutional layer.
Here, $x$ is the input and $o$ is the output data of the layer. 
$C$ and $K$ are, respectively, the number of channels and the width of the kernel.
Each pixel of the data is specified by an index $i$.
The parameters $f$ and $b$ are optimized during the training.
A max pooling layer, 
whose operation can be written as 
\begin{equation}
o^{c}_i = \max_{k=0,1,\cdots,K-1} [x^{c}_{si+k}]\,,
\end{equation}
with the kernel size $K$ and the stride $s$,
reduces the length of the data and hence the computational cost.

In supervised learning, a given dataset consists of many pairs of input data and target values.
An ANN learns the relation between input data and target values from the dataset and predicts 
values corresponding to newly given input data.
In order to train an ANN, the deviation between the predicted 
values and the target value is quantified by a loss function.
For a regression problem, the mean square loss, 
\begin{equation}
L[\ve{y}(w), \ve{t}] = \frac{1}{2} \sum_{i=1}^d |y_i(w) - t_i|^2\,,
\end{equation}
is often employed.
Here, $\ve{y}$ and $\ve{t}$ are a set of predicted values and that of target values, respectively, and they are expressed as $d$-dimensional vectors.
The prediction depends on the weights of the neural network, 
which are denoted by a single symbol $w$.
An ANN is optimized so as to minimize the loss function for a given dataset, 
which is the sum of the loss functions for all data contained in the training dataset.
Because the complete minimization using all dataset cannot be done, the iterative method is used.
The weight $w$ is updated by the replacement algorithm given by 
\begin{equation}
w \to w - \eta \nabla_w \sum_{n=1}^{\Sub{N}{train}} L[\ve{y}_n(w) - \ve{t}_n]\,,
\label{SGD}
\end{equation}
where $\Sub{N}{train}$ is the number of data contained in the dataset and $\eta$ is called 
learning rate and characterizes the strength of each update.
The algorithm shown in Eq.~\eqref{SGD} is called gradient descent, which is the simplest procedure to update the weights,
and many variants (e.g., momentum \cite{momentum}, RMS prop \cite{RMSProp}, Adam \cite{kingma2017adam}) are proposed so far.
Regardless of the choice of the update algorithm, the gradients of a loss function is required and they can be quickly calculated by the backpropagation scheme \cite{BackProp}.
In Eq.~\eqref{SGD}, all data in the dataset are used for each iteration.
In practice, the loss function for a subset of the dataset is calculated.
The subset is called a batch and chosen randomly in every iteration.
This procedure is called a mini-batch training.

In the training process, 
we optimize a neural network so that the loss function is minimized for a dataset.
However, this strategy cannot be straightforwardly applied to practical situations.
First, the trained neural network may fall in overfitting. 
Then, the neural network does not have an expected ability to correctly predict the label for a newly given input data which is not used for training.
Second, we have to optimize the neural network model and the update procedure, too.
For this purpose, we have to appropriately select the hyperparameters, 
such as the number of neurons of the $\ell$-th layer ($N^{(\ell)}$) and the learning rate ($\eta$).
They are not automatically tuned during the training process.

To solve these problems, we prepare a validation dataset which is independent from the training dataset.
The weights of the neural network are optimized so that the loss function for the training dataset is minimized.
The validation data is used for monitoring the training process and assessing which model is better for the problem that the user wants to solve.
To prevent the overfitting, the training should be stopped when the loss for the validation dataset tend to deviate from that for the training dataset (early stopping).
To optimize the hyperparameters, many neural network models having various structures are trained with different training schemes.
Among them, we choose the one performing with the smallest loss for the validation dataset.

\subsubsection{setup in our analysis}

The whole architecture of the neural network we used is shown in Table~\ref{tab:architecture}.
The input data of the neural network is the complex valued numbers taken from a 
short chunk of the $\ell$-domain signal, and the output is the predicted sky position.
The $\ell$-domain waveform $\mathsf{H}_{\ell,k}$ is determined mainly by the residual phase $\delta\Phi_\odot$, which depends on the sky position $(\Sub{\alpha}{s}, \Sub{\delta}{s})$ through the vector $\Delta\ve{n}$.
Because $z_\odot=0$, only $x$ and $y$ components of $\Delta\ve{n}$ affect $\delta\Phi_\odot$.
Therefore, we label each waveform with the values of $\Delta n_x$ and $\Delta n_y$, which are the targets of the prediction of the neural network.
The outputs of the neural network are inverted to the predicted values which are denoted by $(\Sub{\alpha}{p}, \Sub{\delta}{p})$.
We apply the neural network to each candidate, selected by the excess power method, 
in order to narrow down the possible area in which the source is likely to be located.
For simplicity, the $(\alpha,\delta)$-plane is regarded as a two-dimensional Euclidean space, and the shape of the predicted region is assumed to be a disk on the $(\alpha,\delta)$-plane.
For each candidate, the origin of the disk is set to the predicted point.
The radius of the disk, denoted by $\Sub{r}{NN}$, is fixed to a constant value.
In the follow-up stage, the finer grids are placed to cover whole region of the disk.

\begin{table}[t]
\caption{\label{tab:architecture}
The architecture of the neural network used in this work.
For convolution and max pooling layers, the input and the output are characterized by $(C,N)$ where $C$ is the number of channels and $N$ is the length of the data. For convolutional layers, the lengths of kernels are 16, 16, 8, 8, 4 and 4 from the earlier to the later layer. The kernel size of the max pooling layers is 4.}
\begin{ruledtabular}
\begin{tabular}{ccc}
Layer & Input & output \\ \hline
1-d convolution & (2, 2048) &(64, 2033) \\
ReLU & (64, 2033) & (64, 2033) \\
1-d convolution & (64, 2033) & (64, 2018) \\
ReLU & (64, 2018) & (64, 2018) \\
max pooling & (64, 2018) & (64, 504) \\
1-d convolution & (64, 504) & (128, 497) \\
ReLU & (128, 497) & (128, 497) \\
1-d convolution & (128, 497) & (128, 490) \\
ReLU & (128, 490) & (128, 490) \\
max pooling & (128, 490) & (128, 122) \\
1-d convolution & (128, 122) & (256, 119) \\
ReLU & (256, 119) & (256, 119) \\
1-d convolution & (256, 119) & (256, 116) \\
ReLU & (256, 116) & (256, 116) \\
max pooling & (256, 116) & (256, 29) \\
Dense & 256$\times$29 & 64 \\
ReLU & 64 & 64 \\
Dense & 64 & 64 \\
ReLU & 64 & 64\\
Dense & 64 & 2
\end{tabular}
\end{ruledtabular}
\end{table}

In order to train the neural network, 
we need to prepare the training dataset and the validation dataset.
We use Eq.~\eqref{eq:def_of_H} as the model waveform and pick up only a short chunk containing the signal.
The length of chunk is $\delta\ell=2048$.
We prepare 200,000 waveforms for the training and ten thousand waveforms for validation.
At that time, we set $h_0=1$.
We assume that we use only a single detector and use the geometry information (e.g., the latitude of the detector) of LIGO Hanford in calculating the antenna pattern function as an example.
In this work, we focus on one sky patch covered by a single grid point and a frequency bin fixed at $f_k = 100$ Hz since the scaling to the search over the whole sky and the wider frequency band is straightforward.
The sources are randomly distributed within the sky patch.
The parameters $\beta_k$ are randomly sampled from a uniform distribution on $[-0.5, 0.5]$.
The original strain has the duration $2^{24}$ sec and the sampling frequency $1024$ Hz.
We introduce the normalized gravitational wave amplitude by 
\begin{equation}
    \hat{h}_0 := h_0 \left( \frac{\Sub{S}{n}(\Sub{f}{ref})}{\mathrm{1Hz^{-1}}} \right)^{-1/2}.
    \label{eq:normalized_strain}
\end{equation}
Here, we set $\Sub{f}{ref} = f_k$.
At each training step, the amplitude whose logarism is randomly chosen from the uniform distribution on $-2.1 \leq \Sub{\log}{10}\hat{h}_0 \leq -1.0$ is multiplied to the waveforms,
and they are injected into the simulated noise.
The different realizations of noise are sampled for every iterations.
The real part and the imaginary part of the noise data mimicking $\mathsf{N}_{\ell,k}$ are generated from a Gaussian distribution with a zero mean and a variance
\begin{equation}
    \frac{\mathcal{W}}{4\Sub{N}{seg} \Sub{T}{seg}}\,,
\end{equation}
(see Eqs.~\eqref{<NN>} and~\eqref{eq:sigmatilde}).

We employ the mini-batch training. We set the batch size to 256.
The Adam \cite{kingma2017adam} is used for the update algorithm.
We implement with the Python library PyTorch \cite{NEURIPS2019_9015}
and use a GPU GeForce 1080Ti.
The parameter values we used are listed in Table~\ref{tab:params}.
\begin{table}[t]
\caption{\label{tab:params}
The values of the parameters we used in this work.
}
\begin{ruledtabular}
\begin{tabular}{ccc}
Symbol & Parameters & Value \\ \hline
$\Sub{T}{obs}$ & Observation period & $2^{24}$ sec \\
$\Sub{f}{s}$ & Sampling frequency & 1024 Hz \\
$\Sub{N}{grid}$ & \# of grids & 352436 \\
$\Sub{N}{bin}$ & \# of frequency bins of STFT & 3200 \\
$\Sub{T}{seg}$ & Duration of a STFT segment & 32 sec \\
$\Sub{T}{slide}$ & Dilation of STFT segment & 32 sec \\
$\delta\ell$ & Length of chunk & 2048 \\
$\Sub{(\delta\ell)}{slide}$ & Dilation of chunk & 128 \\
$f_k$ & Fixed frequency bin & 100 Hz \\
$\Sub{\alpha}{g}$ & Right ascension of grid & -0.158649 rad\\
$\Sub{\delta}{g}$ & Declination of grid & 1.02631 rad
\end{tabular}
\end{ruledtabular}
\end{table}
%

\subsection{Follow-up analysis by coherent matched filtering}
\label{sec:followup}

After selecting candidates and narrowing down the possible area at which the source 
is likely to be located, we apply 
the coherent matched filtering for the follow-up analysis.
The grid points with the resolution shown in Eq.~\eqref{eq:resolution_coh} are placed to cover the selected area.
Assuming a grid point, we can carry out the demodulation of the phase by using the time resampling technique.
If the deviation between the directions of the grid point and the source is smaller than the resolution, the residual phase remaining after the time resampling is sufficiently small to avoid the loss of SNR.

In this operation, heterodyning and downsampling can significantly reduce the data length and hence the computational cost~\cite{DupuisWoan2005bayesian}.
Let us assume that we have a candidate labeled with $\{\Sub{\ve{n}}{g}, f_k, c\}$.
If the candidate is the true event, the gravitational wave frequency $\Sub{f}{gw}$ should take the value in the narrow frequency band indicated by
\begin{equation}
    f_k - \frac{1}{2\Sub{T}{seg}} \leq \Sub{f}{gw} \leq f_k + \frac{1}{2\Sub{T}{seg}}\,.
    \label{eq:downsampling}
\end{equation}
By multiplying the factor $e^{-2\pi i f_k \zeta}$ to the resampled strain, we can convert the gravitational wave signal frequency to near DC components (\textit{heterodyning}).
After that, the gravitational wave signal has a lower frequency than $1/2\Sub{T}{seg}$ Hz.
Therefore, downsampling by appropriately averaging the resampled strain data with a sampling frequency $\sim 1/\Sub{T}{seg}$ reduces the number of data points without loss of the significance of the gravitational wave signal.

The coherent matched filtering follows the heterodyning and the downsampling processes.
As stated in Eqs.~\eqref{assumption phi0} and~\eqref{assumption psi cosi}, we fix $\psi=0$, $\cos\iota=1$, and $\phi_0=0$ and treat them as known parameters.
Also, we assume that the signal waveform and the template completely match.
The definition of a match is already given in Eq.~\eqref{eq:rho_MF}.
The gravitational waveform is written as
\begin{equation}
    h(\ve{\xi}) = h_0 \cdot \Sub{h}{temp}(\vec{\xi})\,.
\end{equation}
Among these parameters, the amplitude $h_0$ can be analytically marginalized to maximize the likelihood.
Then, we obtain the signal-to-noise ratio in Eq.~\eqref{eq:rho_MF} and use it as the detection statistic.
When only the detector noise dominates the strain data, $\Sub{\rho}{MF}$ follows the standard normal distribution.
On the other hand, if the signal exists, the SNR follows a Gaussian distribution with a mean
\begin{equation}
    \Sub{\mu}{MF}(\ve{\xi}) = h_0 \cdot \sqrt{( \Sub{h}{temp}(\vec{\xi}) | \Sub{h}{temp}(\vec{\xi}))}\,,
\end{equation}
and a unit variance.

\section{Results}
\label{sec:results}

\subsection{Computational cost}
Our procedure is characterized by three parameters $\ve{\rho}:=(\mathrm{FAP}_\mathrm{EP}, \Sub{r}{NN}, \Sub{\hat{\rho}}{MF})$, i.e., 
\begin{itemize}
    \item $\mathrm{FAP}_\mathrm{EP}$, false alarm probability for each chunk,
    \item $\Sub{r}{NN}$, the radius of the predicted region to which the follow-up analysis is applied,
    \item $\Sub{\hat{\rho}}{MF}$, the threshold of the SNR of the coherent matched filtering. 
\end{itemize}

\begin{figure}[t]
    \centering
    \includegraphics[width=8.5cm]{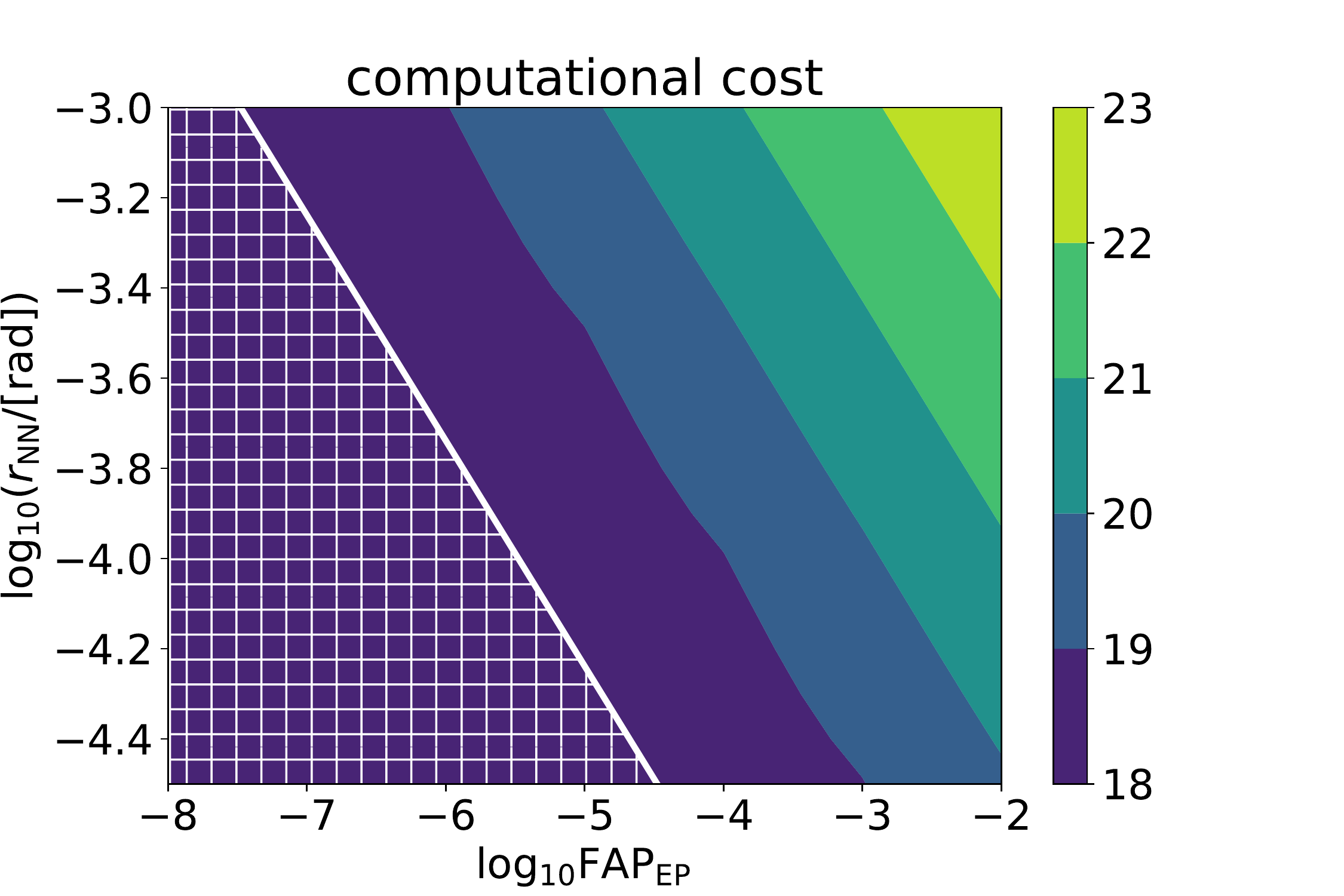}
    \caption{\label{fig:compcost}
    The logarithm of the evaluated computational cost in the unit of the number of floating point operations. In the white hatched region, the computational cost is dominated by that of the preprocessing, i.e., $\Sub{\mathcal{N}}{follow-up} \leq \Sub{\mathcal{N}}{preprocess}$. As the false alarm probability of excess power is set to be smaller, the computational cost is reduced because the number of candidates decreases. Also, the computational cost becomes smaller as the parameter $\Sub{r}{NN}$ is shrunk.}
\end{figure}

$\Sub{\mathcal{N}}{EP}$ denotes the computational cost of the excess power method.
$2\Sub{N}{seg}$ multiplications and $2\Sub{N}{seg}$ additions of real numbers are required to calculate the excess powers for all chunks in one $\ell$-domain signal.
The computational cost for calculating the excess powers for all chunks can be estimated as
\begin{equation}
    \Sub{\mathcal{N}}{EP} = 4\Sub{N}{seg} \times \Sub{N}{grid} \times \Sub{N}{bin} \sim 4.7 \times 10^{15}\,,
\end{equation}
in the unit of the number of floating point operations.
As we see in the following, this cost can be neglected.
Next, we check the computational time of the neural network analysis.
We estimate the computational time of the neural network by measuring the elapsed time for analyzing ten thousand data.
Because the elapsed time is 1.4sec, the total computational time of the neural network is estimated as
\begin{equation}
    \Sub{\mathcal{T}}{NN} \simeq \frac{1.4\mathrm{sec}}{10^4\mathrm{data}} \times \Sub{N}{candidate}\,,
\end{equation}
where $\Sub{N}{candidate}$ is the number of candidates which are selected by the excess power method and is estimated as
\begin{align}
    \Sub{N}{candidate} &= \Sub{N}{chunk} \times \Sub{\mathrm{FAP}}{EP} \notag \\
    &= \Sub{N}{grid} \cdot \Sub{N}{bin} \cdot \frac{\Sub{N}{seg}}{\Sub{(\delta\ell)}{slide}} \times \Sub{\mathrm{FAP}}{EP} \notag \\
    &\simeq 4.6 \times 10^{10} \left( \frac{\mathrm{FAP}_\mathrm{EP}}{10^{-2}} \right)\,.
\end{align}
Substituting it, we obtain
\begin{equation}
    \Sub{\mathcal{T}}{NN} \simeq 6.4 \times 10^6 \mathrm{sec} \left( \frac{\mathrm{FAP}_\mathrm{EP}}{10^{-2}} \right)\,.
\end{equation}
Therefore, we focus on the case $\mathrm{FAP}_\mathrm{EP} \leq 10^{-2}$.
The computational cost of our analysis is dominated mainly by the preprocessing of the observed strain data and the follow-up analysis.
The computational cost of the entire analysis is denoted by $\Sub{\mathcal{N}}{comp}$ and is approximately calculated by
\begin{equation}
    \Sub{\mathcal{N}}{comp} = \Sub{\mathcal{N}}{preprocess} + \Sub{\mathcal{N}}{follow-up}\,,
    \label{eq:Ncomp}
\end{equation}
where $\Sub{\mathcal{N}}{preprocess}$ and $\Sub{\mathcal{N}}{follow-up}$ are the computational cost of the preprocessing and the follow-up analysis, respectively.
In this work, we fix the STFT segment duration and the length of the chunk.
Thus, the computational cost of the preprocessing is a constant.
\begin{equation}
    \Sub{\mathcal{N}}{preprocess} = \Sub{N}{grid} \times ( \Sub{\mathcal{N}}{STFT} + \Sub{\mathcal{N}}{FFT} \times \Sub{N}{bins})\,.
\end{equation}
The computational cost of the STFT is
\begin{equation}
    \Sub{\mathcal{N}}{STFT} = \Sub{N}{seg} \cdot 5 \Sub{T}{seg} \Sub{f}{s} \log_{2} \Sub{T}{seg} \Sub{f}{s}\,,
\end{equation}
and the computational cost of FFT is 
\begin{equation}
    \Sub{\mathcal{N}}{FFT} = 5\Sub{N}{seg} \log_2 \Sub{N}{seg}\,.
\end{equation}
With $\Sub{T}{obs}=2^{24}$sec, $\Sub{T}{seg} = 2^5$sec, $\Sub{f}{s}=2^{10}$Hz, and $\Sub{N}{grid}\sim3.5\times 10^5$, the computational cost of the preprocess is estimated as
\begin{equation}
    \Sub{\mathcal{N}}{preprocess} \simeq 2.3 \times 10^{18}\,.
    \label{eq:Npreprocess num}
\end{equation}
On the other hand, the computational cost of the follow-up analysis is determined by the combination of $\mathrm{FAP}_\mathrm{EP}$ and $\Sub{r}{NN}$.
The computational cost of the follow-up analysis is
\begin{equation}
    \Sub{\mathcal{N}}{follow-up} = \Sub{N}{candidate} \times \frac{\pi (\Sub{r}{NN})^2}{\Sub{(\delta\theta)}{coh}^2} \times \Sub{\mathcal{N}}{FFT,coh}\,.
\end{equation}
Here, $\Sub{(\delta\theta)}{coh}^2$ is the typical area of region where each grid point of the coherent analysis covers (see Eq.~\eqref{eq:resolution_coh}).
The computational cost of taking match is dominated by the Fourier transform and calculated as
\begin{equation}
    \Sub{\mathcal{N}}{FFT,coh} = 5 (\Sub{T}{obs} \Sub{f}{s,coh}) \log_2 (\Sub{T}{obs} \Sub{f}{s,coh}) \simeq 5.0 \times 10^7\,,
\end{equation}
where $\Sub{f}{s,coh} = 1/\Sub{T}{seg} = 2^{-5}$Hz.
Therefore, we estimate the computational cost of the follow-up analysis as
\begin{equation}
    \Sub{\mathcal{N}}{follow-up} = 9.0 \times 10^{22} \left( \frac{\Sub{r}{NN}}{10^{-3}\mathrm{rad}} \right)^2 \left( \frac{\mathrm{FAP}_\mathrm{EP}}{10^{-2}} \right)\,.
    \label{eq:Nfolloup num}
\end{equation}
Substituting Eqs.~\eqref{eq:Npreprocess num} and~\eqref{eq:Nfolloup num} into Eq.~\eqref{eq:Ncomp}, we can assess the computational cost of the entire analysis as a function of $\Sub{r}{NN}$ and $\mathrm{FAP}_\mathrm{EP}$.
Figure.~\ref{fig:compcost} shows the computational cost for various combinations of $\mathrm{FAP}_\mathrm{EP}$ and $\Sub{r}{NN}$.
One can read a feasible combination of $\mathrm{FAP}_\mathrm{EP}$ and $\Sub{r}{NN}$ depending on one's available computational resources.

\subsection{False alarm probability}

The false alarm probability of the entire process (see \cite{Wette2011, Dreissigacker:2018afk}) is 
\begin{align}
    \Sub{p}{fa}(\ve{\rho}) = &\left\{ 1 - \left(\mathrm{Prob}\left[ \Sub{\rho}{EP} < \Sub{\hat{\rho}}{EP} \mid \Sub{\rho}{EP} \sim \mathcal{N}(0,1) \right] \right)^{\Sub{N}{chunk}} \right\}\notag \\
    &\times \left\{ 1 - \left(\mathrm{Prob}\left[ \Sub{\rho}{MF} < \Sub{\hat{\rho}}{MF} \mid \Sub{\rho}{MF} \sim \mathcal{N}(0,1) \right]\right)^{\Sub{N}{t}} \right\}\,,
    \label{eq:totalFAP}
\end{align}
where $\Sub{N}{t}$ is the number of required templates for the coherent search.
It can be estimated by
\begin{equation}
    \Sub{N}{t} = \Sub{N}{candidate} \times \frac{\pi (\Sub{r}{NN})^2}{(\delta\theta^2)_\mathrm{coh}} \times \Sub{N}{bin,coh},
\end{equation}
where $\Sub{N}{bin,coh}$ is the number of the frequency bins of the coherent search.
Using the value listed in Table.~\ref{tab:params}, we obtain
\begin{equation}
    \Sub{N}{t} \simeq 5.6 \times 10^{21} \left( \frac{\Sub{r}{NN}}{10^{-3}\mathrm{rad}} \right)^2 \left( \frac{\mathrm{FAP}_\mathrm{EP}}{10^{-2}} \right)\,.
    \label{Eq:Ntemp}
\end{equation}
Because the false alarm probability of the follow-up stage determines that of the entire process, we can approximate it as
\begin{equation}
    \Sub{p}{fa}(\ve{\rho}) \simeq \left\{ 1 - \left(\mathrm{Prob}\left[ \Sub{\rho}{MF} < \Sub{\hat{\rho}}{MF} \mid \Sub{\rho}{MF} \sim \mathcal{N}(0,1) \right]\right)^{\Sub{N}{t}} \right\}\,.
\end{equation}
Furthermore, because $\Sub{N}{t} \gg 1$, we can approximate
\begin{equation}
    \Sub{p}{fa}(\ve{\rho}) \simeq \Sub{N}{t} \cdot \mathrm{Prob}\left[ \Sub{\rho}{MF} > \Sub{\hat{\rho}}{MF} \mid \Sub{\rho}{MF} \sim \mathcal{N}(0,1) \right]\,.
\end{equation}
In this work, the threshold $\Sub{\hat{\rho}}{MF}$ is chosen so that the false alarm probability of the entire process is 0.01.
As shown in Eq.~\eqref{Eq:Ntemp}, the number of templates depends on $(\Sub{r}{NN}, \mathrm{FAP}_\mathrm{EP})$, and the same is true for $\Sub{\hat{\rho}}{MF}$.

\begin{figure}[t]
    \centering
    \includegraphics[width=9cm]{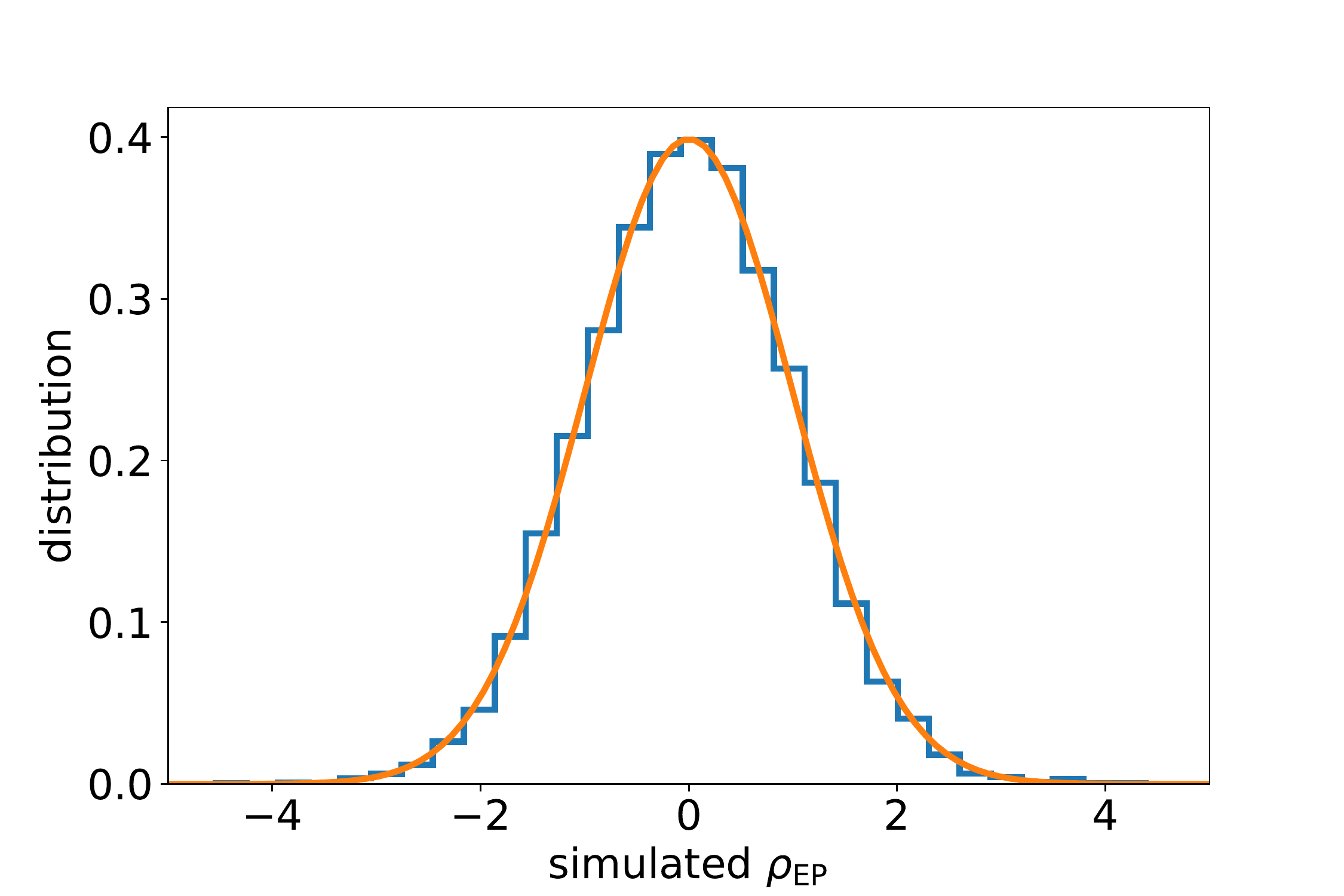}
    \caption{\label{fig:SNREP_noise}
    The histogram of the simulated $\Sub{\rho}{EP}$.
    The blue line is the histogram, and the orange line indicates the standard normal distribution. They match well. The p-value of the Kolmogorov-Smirnov test is 0.753254.}
\end{figure}

The false alarm probability of the matched filtering has already been studied in the literature.
Therefore, in this work, we check only the validity of the statistical properties of the excess power method.
It is computationally difficult to treat a whole signal of a duration $\Sub{T}{obs} = 2^{24}$sec.
Therefore, we generate $\Sub{N}{seg}$ short noise data of a duration $\Sub{T}{seg}$ assuming $\Sub{S}{n}(f) = 1$.
After applying a window function, FFT is carried out to each short strain.
We pick up a FFT of $k$-th frequency bin from each FFT data and regard them as $\{\Super{n}{STFT}_{j,k}\}_{j=1}^{\Sub{N}{seg}}$.
We obtain $\mathsf{N}_{\ell,k}$ by taking the Fourier transform of $\{\Super{n}{STFT}_{j,k}\}_{j=1}^{\Sub{N}{seg}}$ and divide it into $\Sub{N}{seg} / \delta\ell = 128$ chunks.
After repeating above procedures for 80 times, 10,240 chunks are generated.
For each chunk, the excess power statistics $\mathcal{E}$ and SNRs $\Sub{\rho}{EP}$ are calculated.
The histogram of the simulated values of $\Sub{\rho}{EP}$ is shown in Fig.~\ref{fig:SNREP_noise}.
It seems to match the standard normal distribution.
Additionally, we carry out the Kolmogorov-Smirnov test and obtain a p-value of 0.753254.
It is numerically confirmed that the SNR of noise data follows the standard normal distribution.

\subsection{Detection probability}
The detection probability of the signal with an amplitude $h_0$ can be estimated by
\begin{align}
    &\ \Sub{p}{det}(h_0; \ve{\rho}) \notag \\
    &= \left\langle \Sub{p}{det}(\ve{\xi}; \ve{\rho}) \right\rangle_{\vec{\xi}} \notag \\
    =& \bigg\langle \mathrm{Prob}\left[ \Sub{\rho}{EP} > \Sub{\hat{\rho}}{EP} \mid \Sub{\rho}{EP} \sim \mathcal{N}(\Sub{\mu}{EP}(\ve{\xi}), \Sub{\sigma}{EP}(\ve{\xi})) \right] \notag \\
    &\times \mathrm{Prob}\left[ \text{source is located in a predicted region} \mid \ve{\xi}, \Sub{r}{NN} \right] \notag \\
    &\times \mathrm{Prob}\left[ \Sub{\rho}{MF} > \Sub{\hat{\rho}}{MF} \mid \Sub{\rho}{MF} \sim \mathcal{N}(\Sub{\mu}{MF}(\ve{\xi}), 1 ) \right] \bigg\rangle_{\vec{\xi}}\,,
    \label{eq:totalFDPexact}
\end{align}
where
\begin{equation}
    \langle \cdots \rangle_{\vec{\xi}} := \int d\vec{\xi}\ (\cdots) \pi(\vec{\xi})\,,
\end{equation}
is the average over the source parameters $\vec{\xi}$ with the probability density function $\pi(\vec{\xi})$.
As explained in Sec.~\ref{sec:NN}, the neural network is trained with the waveforms sampled only from the vicinity of the reference grid point and the narrow frequency band.
It is envisaged that the trained neural network does not work well for signals outside of the reference patch and the frequency band.
Therefore, we only test for the limited parameter region.
Correspondingly, the average operation is also taken over such narrow parameter space.
To quantify the detection power, the amplitude parameter $\Super{h}{95\%}_0$ is defined by
\begin{equation}
    \Sub{p}{det}(\Super{h}{95\%}_0; \ve{\rho}) = 0.95\,,
\end{equation}
and correspondingly,
\begin{equation}
    \Super{\hat{h}}{95\%}_0 := \Super{h}{95\%}_0 \left( \frac{\Sub{S}{n}(\Sub{f}{ref})}{\mathrm{1Hz^{-1}}} \right)^{-1/2}.
\end{equation}
The parameters, $\mathrm{FAP}_\mathrm{EP}$ and $\Sub{r}{NN}$, are optimized so that $\Super{h}{95\%}_0$ takes the smallest value under the condition of the computational power.

\begin{figure}[t]
    \centering
    \includegraphics[width=8.5cm]{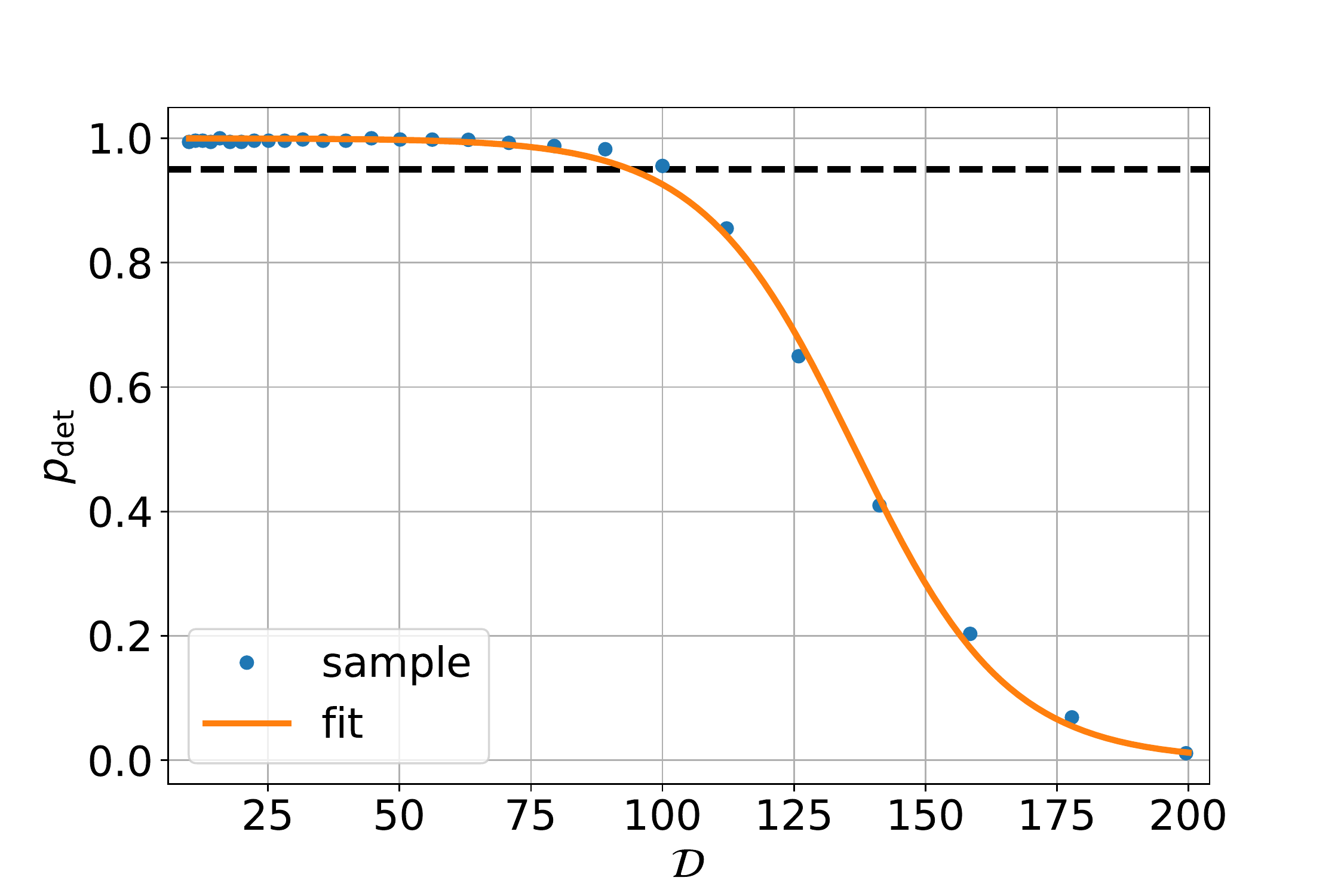}
    \caption{\label{fig:fitexample}
    Example of fitting of detection probability.
    We set $\mathrm{FAP}_\mathrm{EP} = 10^{-3}$ and $\Sub{r}{NN} = 10^{-3.8}$rad.
    Blue dots are estimated values of $\Sub{p}{det}$, and orange solid line is the fitted sigmoid curve.}
\end{figure}

\begin{figure}[t]
    \centering
    \includegraphics[width=8.6cm]{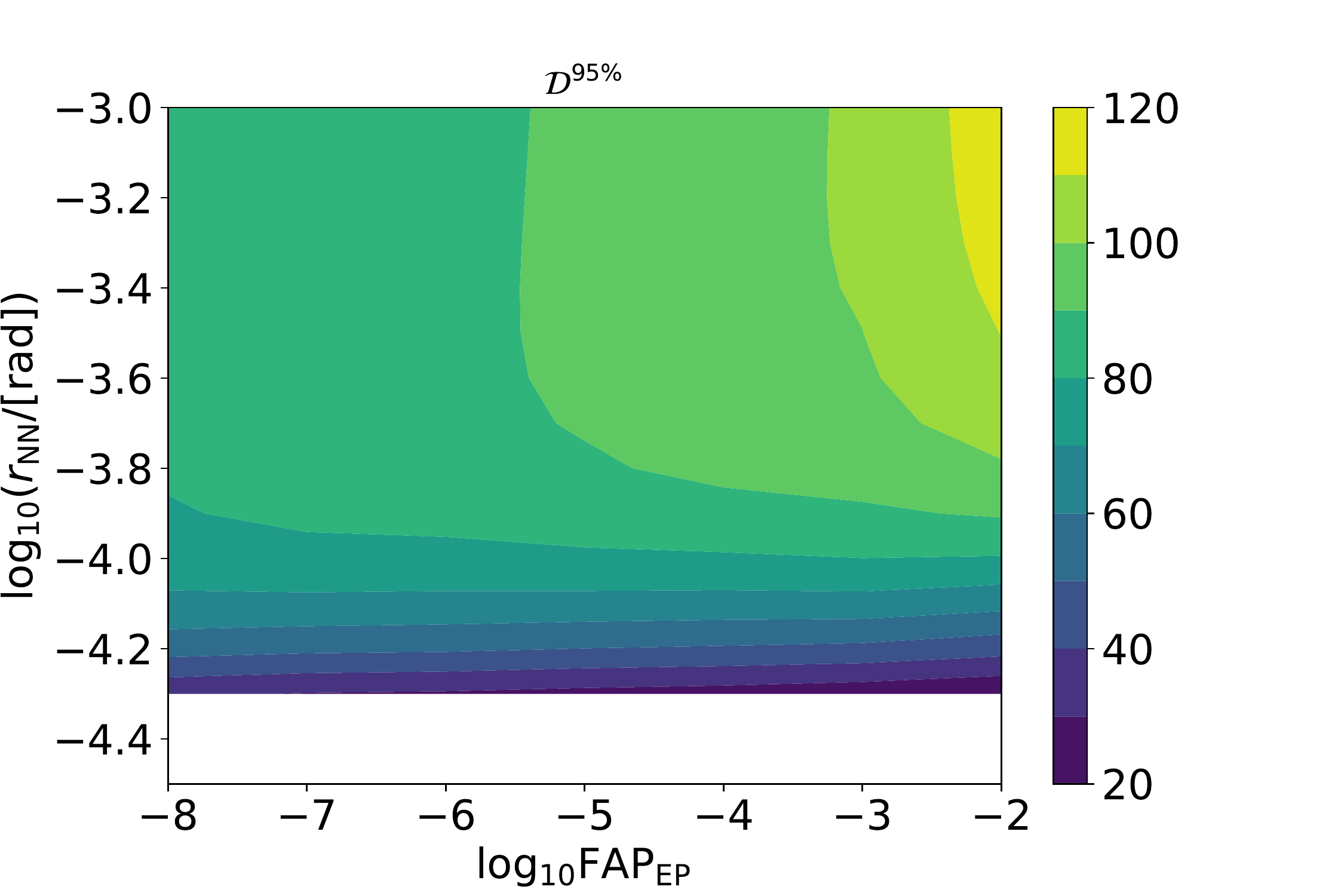}
    \caption{\label{fig:D95est}
    Estimated sensitivity depths $\Super{\mathcal{D}}{95\%}$.
    For parameters within a blank region, the detection probabilities do not reach 95\% for a surveyed range of amplitude $\hat{h}_0$.
    For $\Sub{r}{NN}\lesssim 10^{-4}\mathrm{rad}$, the neural network controls the detection probability. On the other hand, for $\Sub{r}{NN} \gtrsim 10^{-3.6}\mathrm{rad}$, the excess power method controls the detection probability because the predicted region is large enough to contain the true location of the source.}
\end{figure}

To explorer the parameter space of ($\mathrm{FAP}_\mathrm{EP}, \Sub{r}{NN}$), we place the regular grid on $\log_{10}\mathrm{FAP}_\mathrm{EP}$ from $-8$ to $-2$ by a step of 1, and the regular grid on $\log_{10} \Sub{r}{NN}$ from -4.5 to -3.0 by a step of 0.05.
For every pair of $\mathrm{FAP}_\mathrm{EP}$ and $\Sub{r}{NN}$, we calculate $\Super{\hat{h}}{95\%}_0$ by the following procedure.
First, we place a regular grid on $\log_{10} \hat{h}_0$ from -2.3 to -1.0 by a step of 0.05.
For one sample of the amplitude, the parameters $\vec{\xi}$ are randomly sampled.
The sampled parameters are denoted by $\{ \vec{\xi}^{(i)} \}_{i=1}^{M}$.
The waveforms are generated with the sampled parameters.
Each waveform is injected into different noise data in the same manner as the method explained in Sec.~\ref{sec:NN}.
The fraction of the events detected is employed as the estimator of the detection probability of the signal with parameter $(h_0, \vec{\xi}^{(i)})$.
Repeating these procedure for every sampled parameters $\{ \vec{\xi}^{(i)} \}_{i=1}^{M}$, we obtain the set of the estimated detection probabilities.
Then, the detection probability $\Sub{p}{det}(h_0; \ve{\rho})$ is estimated by
\begin{equation}
    \Sub{p}{det}(h_0; \ve{\rho}) \simeq \frac{1}{M} \sum_{i=1}^M \Sub{p}{det}(h_0, \vec{\xi}^{(i)}; \ve{\rho})\,.
\end{equation}
Changing the value of the amplitude $\hat{h}_0$, we get the estimated detection probability as a function of $h_0$ for a certain values of $\mathrm{FAP}_\mathrm{EP}$ and $\Sub{r}{NN}$.
If the estimated detection probability exceeds 95\% for one or more samples of the amplitude, the obtained detection probabilities are fitted by a sigmoid-like function,
\begin{equation}
    \varsigma(\mathcal{D}; a, b) = \frac{1}{1 + e^{(\mathcal{D}-a)/b}}\,,
\end{equation}
where 
\begin{equation}
    \mathcal{D}:=(\hat{h}_0)^{-1}\,,
\end{equation}
is called \textit{the sensitivity depth}, and the parameters $(a,b)$ of a sigmoid function is to be optimized.
Using the optimized parameters $(a^\ast, b^\ast)$, the estimated value of $\Super{\mathcal{D}}{95\%}:=(\Super{\hat{h}}{95\%}_0)^{-1}$ can be obtained as 
\begin{equation}
    \Super{\mathcal{D}}{95\%}(\mathrm{FAP}_\mathrm{EP}, \Sub{r}{NN}) = a^\ast - b^\ast \ln \frac{1 - 0.95}{0.95}\,,
\end{equation}
which is the inverse of $\varsigma(\mathcal{D})$.
In this work, we set $M=1024$ and the number of noise realization for each parameter set to be 512.
Figure.~\ref{fig:fitexample} shows an example of the fitting.
The estimated values of $\Super{\mathcal{D}}{95\%}$ is shown in Fig.~\ref{fig:D95est}.


\begin{figure}[t]
    \centering
    \includegraphics[width=8.5cm]{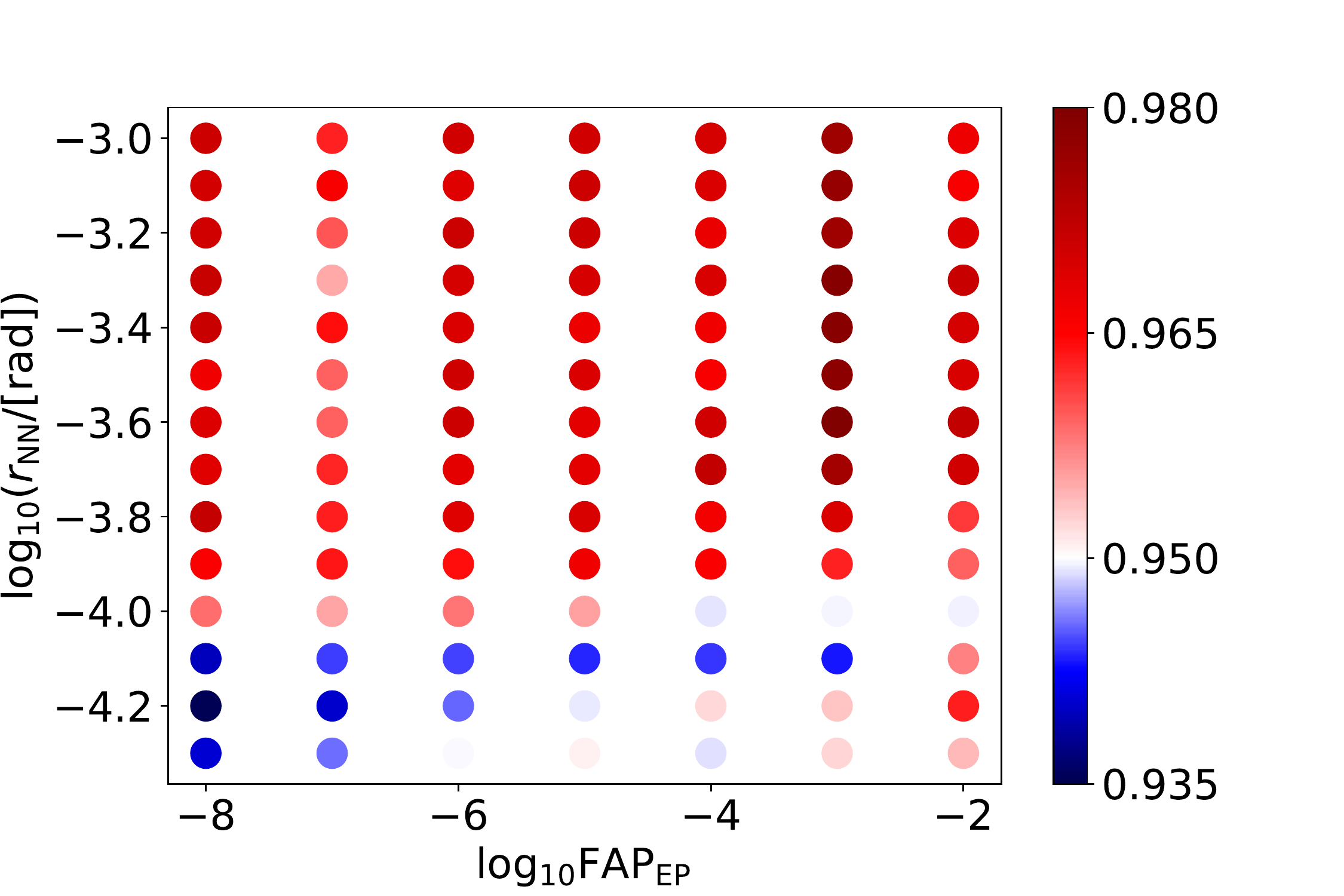}
    \caption{\label{fig:recovered pdet}
    Recovered values of the detection probability.
    For all parameters, the detection probabilities are recovered to 95\% with the error of only a few percent.}
\end{figure}

To confirm that the signals with the amplitude $\Super{h}{95\%}_0$ are detected with 95\% detection probability, we perform the injection test.
To save the computational cost, we skip the follow-up stage and assume that the detection probability is determined by the excess power selection and the neural network analysis.
We only use a short chunk centered at the support of the signal as an injection waveform.
Ten thousand chunks with various signal parameters $\vec{\xi}$ are prepared and injected into Gaussian noise data.
The waveform model and the noise property are the same as those of the training dataset of the neural network.
The excess power is calculated for each chunk, and the neural network analysis is carried out if a chunk is selected as a candidate.
Counting the number of detected events, we obtain the recovered value of the detection probability.
The procedure shown above is repeated for each combination of $(\mathrm{FAP}_\mathrm{EP}, \Sub{r}{NN})$.
Figure.~\ref{fig:recovered pdet} shows the result of the injection test.
For all combinations of $(\mathrm{FAP}_\mathrm{EP}, \Sub{r}{NN})$, the detection probabilities are close to 95\%.
Therefore, our estimation of the detectable amplitude $\Super{\hat{h}}{95\%}_0$ is convincing.

\section{Conclusion}
\label{sec:conclusion}

We proposed a new method of an all-sky search for continuous gravitational waves, combining the excess power and the deep learning methods.
The time resampling and the STFT are used for localizing the signal into a relatively small number of elements in the whole data.
Then, the excess power method selects the candidates of the grid point in the sky and the frequency bins where the signal likely exists.
The deep neural network narrows down the region to be explored by the follow-up search by two orders of magnitude than the original area of the sky patch.
Before the follow-up coherent search, the heterodyning and the downsampling can reduce the computational cost.
We calculated the computational cost of our method.
Most of the computational costs are spent by preprocessing the strain data and the follow-up coherent matched filtering search.
The computational costs of the excess power method are negligibly small, and the computational time of the neural network can be suppressed to an acceptable level by setting $\mathrm{FAP}_\mathrm{EP}\leq 10^{-2}$.
We estimated the detection abilities of our method with the limited setup where the polarization angle, the inclination angle, and the initial phase are fixed and assumed as known parameters.
The dataset for training the neural network and testing is generated from a very narrow parameter space of $(\Sub{f}{gw}, \Sub{\alpha}{s}, \Sub{\delta}{s})$.
With a reasonable computational power, the sensitivity depth can be achieved $\Super{\mathcal{D}}{95\%} \gtrsim 80$.

Our training data, which is used for training the neural network, span the restricted parameter region.
Namely, the gravitational wave frequencies of the training data are distributed within the small frequency band centered at 100 Hz of width $\pm 1/(2\Sub{T}{seg})$ and the source locations are sampled from very narrow regions around the fixed grid point.
Nevertheless, we can expect our method can be applied to the all-sky search and the frequency band below 100 Hz.
If the gravitational wave frequency becomes lower than 100Hz, the strength of the phase modulation becomes weaker (see Eq.~\eqref{eq:modulatedphase}).
Therefore, even if $\Sub{f}{gw}<100$ Hz, the signal power in $\ell$-domain would still be concentrated in a narrow region, and it can be expected that the efficiency of the excess power method is maintained.
We can employ a similar discussion also for the dependency on the source location.
The power concentration in $\ell$-domain is still valid even if we take into account the dependency of the source location, while it causes the variation of the signal amplitude.
From the above discussion, only slight modifications of the construction of the training data and our neural network structure are enough to apply our strategies to an all-sky search of monochromatic sources having a frequency lower than $\sim 100$ Hz.

In addition to the above points, there are several rooms for improving our method.
We fixed various parameters such as the width of the STFT $\Sub{T}{seg}$ and the length of each chunk $\delta\ell$ in a little hand-waving manner.
Surveying and optimizing these parameters may improve the detection efficiency of our method.
Especially, the sampling frequency when downsampling might reduce the computational cost significantly.
As can be seen from Eq.~\eqref{eq:ell_domain_signal}, the deviation $\delta f_k$ causes the translation of the signal in the $\ell$-domain.
It is expected that we can further constrain the gravitational wave frequency than $\sim \Sub{T}{seg}^{-1}$.
Considering this effect, we can set the sampling frequencies of downsampled strains to a lower value than our current choice.
This optimization would result in the further reduction of the computational time of the follow-up coherent search.

In the present paper, we assumed that the stationary Gaussian detector noise and 100\% duty cycle.
We also simplified the waveform model, e.g., the frequency change $df/dt$ is not incorporated.
In spite of these simplifications,  
the obtained results can be regarded as a proof-of-principle and are enough to convince that our method has the potential for improving the all-sky search for continuous gravitational waves with the duration of $O(10^7)$ sec.
Relaxing these assumptions is beyond the scope of this paper and left as future work.


\begin{acknowledgements}

We thank Yousuke Itoh for fruitful discussions. This work was supported by JSPS KAKENHI Grant Number JP17H06358 (and also JP17H06357), \textit{A01: Testing gravity theories using gravitational waves}, as a part of the innovative research area, ``Gravitational wave physics and astronomy: Genesis'', and JP20K03928. Some part of calculation has been performed by using a GeForce 1080Ti GPU at Nagaoka University of Technology.

\end{acknowledgements}

\appendix
\section{Noise statistics in $\ell$-domain}
\label{sec:appendix}

In general, the power spectral density of a stochastic process $n(t)$ is defined by
\begin{equation}
    \langle \tilde{n}(f) \tilde{n}^\ast(f') \rangle =: \frac{1}{2}\Sub{S}{n}(f) \delta(f-f')\,,
\end{equation}
where the Fourier transform of $n(t)$ is defined by
\begin{equation}
    \tilde{n}(f) = \int^{\infty}_{-\infty} dt\ n(t) e^{-2\pi i ft}\,,
\end{equation}
while we define the STFT by Eq.~\eqref{eq:n_STFT}.
Ignoring the effect of the window function, the variance of $\Super{n}{STFT}_{j,k}$ can be approximated by
\begin{equation}
    \langle (\Super{n}{STFT}_{j,k}) (\Super{n}{STFT}_{j',k'})^\ast \rangle = \frac{1}{2\Sub{T}{seg}} \Sub{S}{n}(f_k)  \delta_{kk'} \delta_{jj'}\,.
\end{equation}
Here, we assume that different STFT bins are statistically independent.
The variance of $\mathsf{N}_{\ell,k}$ is 
\begin{align}
    &\langle \mathsf{N}_{\ell,k} \mathsf{N}^\ast_{\ell',k'} \rangle \notag\\
    &= \frac{1}{\Sub{N}{seg}^2} \sum_{j=1}^{\Sub{N}{seg}} \sum_{j'=1}^{\Sub{N}{seg}} \langle (\Super{n}{STFT}_{j,k}) (\Super{n}{STFT}_{j',k'})^\ast  \rangle e^{-2\pi i (j\ell - j'\ell' )/\Sub{N}{seg}} \notag\\
    &= \frac{\Sub{S}{n}(f_k)}{2\Sub{T}{seg}\Sub{N}{seg}} \delta_{\ell \ell'} \delta_{k k'}\,.
\end{align}
Therefore, we get
\begin{equation}
    \tilde{\sigma}_k^2 = 2 \langle \mathsf{N}_{\ell,k} \mathsf{N}^\ast_{\ell, k} \rangle = \frac{\Sub{S}{n}(f_k)}{\Sub{T}{seg}\Sub{N}{seg}}\,.
\end{equation}


\bibliography{references, refs_dl}

\end{document}